\documentclass[12pt]{elsart}

\usepackage{amsmath}
\usepackage{amssymb}
\usepackage{amsfonts}
\usepackage{graphicx}
\usepackage{epstopdf}
\usepackage{color}
\usepackage{bm}

\usepackage{ifpdf}
\ifpdf%
\usepackage{pdflscape}
\else
\usepackage{lscape}
\fi

\DeclareMathOperator{\Tr}{Tr}
\DeclareMathOperator{\tr}{tr}
 
\DeclareMathOperator{\Imag}{Im}

\begin{document}

\begin{frontmatter}
\title{Wave function multifractality and dephasing at metal-insulator
  and quantum Hall transitions}
\author[ISB1,ISB2]{I.S. Burmistrov\corauthref{lc1}}
\corauth[lc1]{Corresponding author. Fax: +7-495-7029317 } \ead{burmi@itp.ac.ru}
\author[IVG1,ADM1]{S. Bera}
\author[IVG1,ADM1]{  F. Evers}
\author[IVG1,IVG2]{  I.V. Gornyi}
\author[IVG1,ADM1,ADM2]{A.D. Mirlin}

\address[ISB1]{L.D. Landau Institute for Theoretical Physics,
Kosygina street 2, 117940 Moscow, Russia}
\address[ISB2]{Department
of Theoretical Physics, Moscow Institute of Physics and
Technology, 141700 Moscow, Russia}
\address[IVG1]{Institut f\"ur Nanotechnologie, Karlsruhe Institute of
  Technology, 76021 Karlsruhe, Germany}
\address[ADM1]{Institut f\"ur Theorie der kondensierten Materie,
  Karlsruhe Institute of Technology, 76128 Karlsruhe, Germany}
\address[IVG2]{A. F. Ioffe Physico-Technical Institute, 194021
  St. Petersburg, Russia}
\address[ADM2]{Petersburg Nuclear Physics Institute, 188300
  St. Petersburg, Russia}

\begin{abstract}
We analyze the critical behavior of the dephasing rate induced by short-range
electron-electron interaction near an Anderson transition of
metal-insulator or quantum Hall type. The corresponding exponent
characterizes the scaling of the transition width with temperature.
Assuming no spin degeneracy, the critical behavior can be studied by
performing the scaling analysis in the vicinity of the non-interacting
fixed point, since the latter  is stable with respect to the
interaction. We combine an analytical treatment (that includes the
identification of operators responsible for dephasing in the
formalism of the non-linear sigma-model and the corresponding
renormalization-group analysis in $2+\epsilon$ dimensions) with
numerical simulations on the Chalker-Coddington network model of the
quantum Hall transition. Finally, we discuss the current understanding
of the Coulomb interaction case and the available experimental data.

\end{abstract}


\end{frontmatter}

%%%%%%%%%%%%%%%%%%%%%%%%%%%%%%%%%%%%%%%%%%%%%%%%%%%%%%%%%%%%%%

\section{Introduction}
\label{intro}

Localization-delocalization quantum phase transitions
form a broad and actively developing field of condensed
matter physics, see Ref.~\cite{evers08} for a recent review. In
particular, the electron-electron interaction effects at the
transition represent one of central research directions
\cite{AA85,finkelstein90,belitz94,pruisken10,abrahams01,pudalov04}.

Quite generally, the impact of interaction onto low-temperature
transport and localization in disordered electronic systems can be
subdivided into effects of (i) renormalization and (ii) dephasing.
The renormalization effects, whose role in diffusive systems was
investigated by Altshuler and Aronov \cite{AA85}, are governed by
virtual processes and become increasingly more pronounced with
lowering temperature. Finkelstein developed a renormalization-group
(RG) approach \cite{finkelstein90} in the framework of the non-linear
$\sigma$-model in order to treat these effects together with
localization phenomena. More recently, this research direction
attracted a great deal of attention in connection with experiments on
high-mobility low-density electronic structures (Si MOSFETs)
giving an evidence in favor of a metal-insulator 
transition \cite{abrahams01,pudalov04,2DMIT}. It was shown
that this transition  can be explained in the framework of the
$\sigma$-model RG for a system with $N > 1$ valleys \cite{punnoose05}
(formally $N$ should be large but in practice $N=2$, as in Si, is already
sufficient). Indeed, a detailed analysis has confirmed that the RG
theory describes well the experimental data up to lowest accessible
temperatures \cite{anissimova07,knyazev08}. Very recently, it was shown
\cite{ostrovsky10} that interaction effects are also of crucial
importance in topological insulators, where they induce novel critical
states. Another long-standing issue in the field concerns the 
interplay of interaction and multifractality of wave functions at Anderson 
transitions. This problem has been recently addressed
in the context of superconductivity near the mobility 
edge. Focussing on the short-range interaction in the Cooper channel, Ref. 
\cite{feigelman10} predicted that the multifractality of critical wave 
functions strongly enhances  the superconducting pairing correlations.

Dephasing effects are governed by inelastic processes of
electron-electron scattering at finite temperature $T$. The dephasing
has been studied in great detail for metallic systems where it
provides a cutoff for weak-localization effects \cite{AA85}. As
to Anderson localization transitions, dephasing leads to their
smearing at finite temperature. The dephasing-induced transition width
scales as a power-law function of $T$. In the case of quantum Hall
transition, this scaling of the transition width, $\Delta B \propto
T^\kappa$, has been experimentally explored in many works. While the
value of $\kappa$ was a matter of controversy, there seems to be a
consensus now that the results $\kappa=0.42 \pm 0.04$ \cite{wei88},
 $\kappa=0.42 \pm 0.01$ \cite{li05} properly characterize the quantum
 Hall transition (when it is not masked by macroscopic
 inhomogeneities). Scaling near the transition with varying temperature
was also experimentally studied at the 3D Anderson transition in doped
semiconductors \cite{waffenschmidt99,bogdanovich99}.
% but the scaling analysis in these systems is more complicated,
% leading to considerably larger uncertainties in experimental values
% of the critical exponents.

In this paper we study the scaling properties of dephasing rate at
criticality. We consider the situation when the time-reversal invariance
is broken, so that the system belongs to the so-called unitary symmetry
class. In particular, the quantum Hall transition belongs to this
symmetry class. We further assume, following
Refs.~\cite{LeeWang,Wang2} that the interaction is of short-range
character. This greatly simplifies the analysis, since in this case
the interaction is irrelevant in the RG sense \cite{belitz94,LeeWang,PB}, so that all scaling
properties at the transition are governed by the RG flow near the
non-interacting fixed point. The goals of this work are as
follows. First, we identify the operators controlling the dephasing
within the framework of the non-linear $\sigma$-model.
Second, we perform the RG analysis of these operators for
the $\sigma$-model in $2+\epsilon$ dimensions. Third, we
present a numerical investigation of the corresponding wave function
correlation functions by using the Chalker-Coddington network as a
a model of the quantum Hall transition. Finally, we make conclusions
concerning the exponent $\kappa$ governing the temperature scaling of
the transition and discuss the current understanding of the Coulomb
interaction case, as well as relations between the theoretical and
experimental findings.

\section{Dephasing rate and exact eigenfunction of the
non-interacting   problem}
\label{s1}

\subsection{Model}
\label{s1.1}

We consider a disordered interacting electronic system characterized
by the Hamiltonian
\begin{equation}
\label{e1}
\mathcal{H} = \mathcal{H}_0+\mathcal{H}_{\rm int}\,.
\end{equation}
The one-particle term in the Hamiltonian,
\begin{equation}
\label{e2}
\mathcal{H}_0 = \int d\bm{r}\,
\Psi^{\dag}(\bm{r})\Bigl [
\frac{1}{2m_e}(-i\nabla - \bm{A}(\bm{r}))^2+V_{\rm dis}(\bm{r})\Bigr
] \Psi(\bm{r})\,,
\end{equation}
is assumed to describe a system with broken time-reversal invariance
(unitary symmetry class) at criticality. This may be a 2D system near
quantum Hall transition (in which case $\bm{B} = {\rm rot}\bm{A}$ is a
uniform magnetic field), or a system in $d>2$ dimensions at the
Anderson transition point (with a uniform or random magnetic field
$\bm{B}(\bm{r})$ breaking the time reversal invariance).
 The interaction part of the Hamiltonian $\mathcal{H}$ has the form
\begin{equation}
\label{e3}
\mathcal{H}_\textrm{int} =\,  \frac{1}{2}\int
d\bm{r_1} d\bm{r_2}\,U(\bm{r}_1-\bm{r}_2) \Psi^\dag(\bm{r}_1)
 \Psi^\dag(\bm{r}_2) \Psi(\bm{r}_2) \Psi(\bm{r}_1) \,,
\end{equation}
with a short-range interaction potential $U(r)$.

\subsection{First-order correction to the thermodynamic potential}
\label{s1.2}

%%%%%%%%%%%%%%%%%%%%%%%%%%%%%%%%%%%%%%%%%%%%%%%%%%%%%%%%%%%%%%%%%%%%%%%%%%%%%
\begin{figure}[t]
\centerline{\includegraphics[width=100mm]{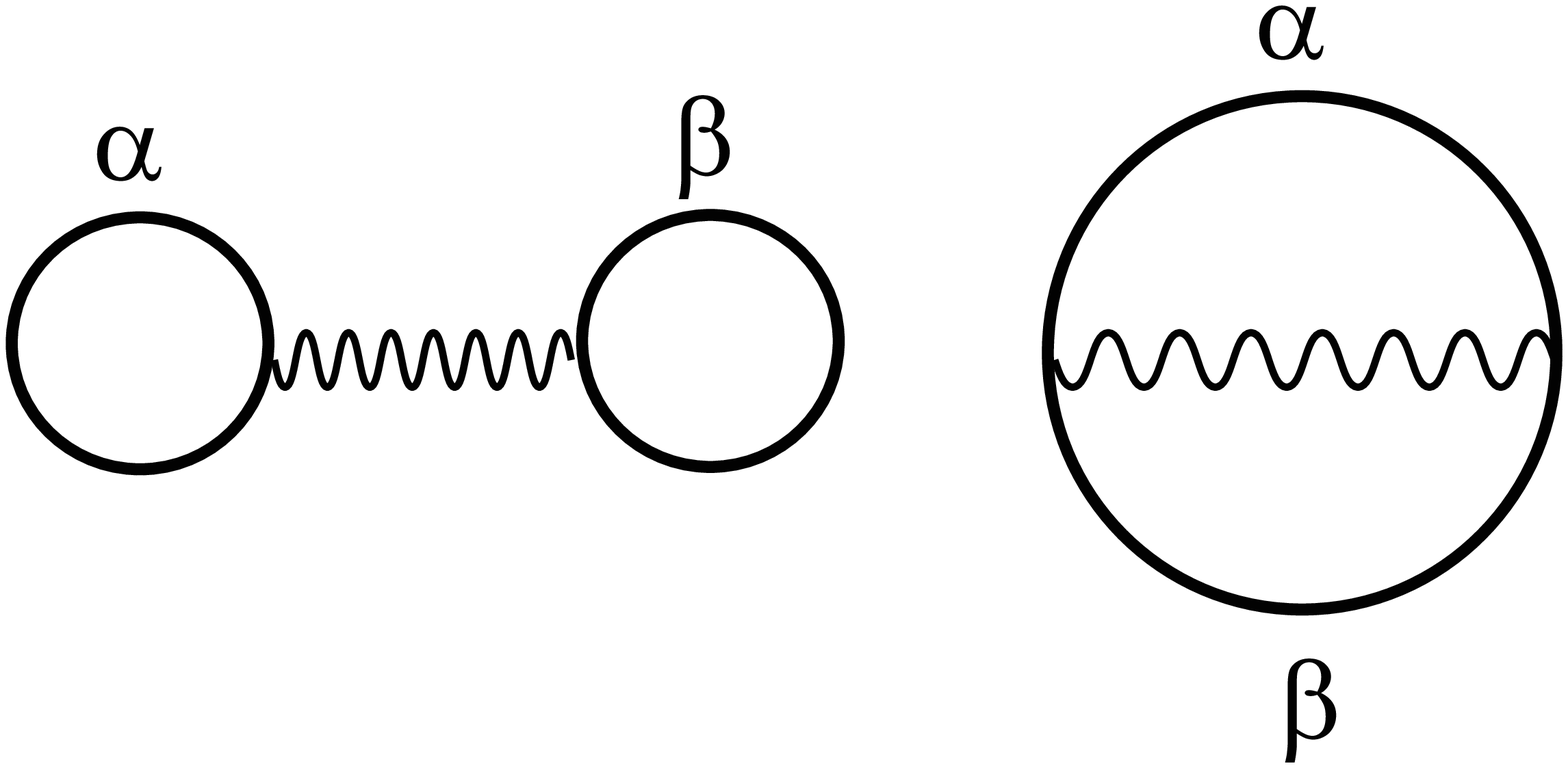}}
\caption{The first-order (Hartree and Fock)
interaction corrections to the thermodynamic
  potential. The wavy line denotes the interaction and the solid line
  the electron Green's function.} \label{Figure1}
\end{figure}
%%%%%%%%%%%%%%%%%%%%%%%%%%%%%%%%%%%%%%%%%%%%%%%%%%%%%%%%%%%%%%%%%%%%%%%%%%%%%

Following Ref.~\cite{LeeWang}, we begin by considering
the first order interaction correction to the thermodynamic potential.
The corresponding diagrams are shown
in Fig.~\ref{Figure1}; their total contribution reads
\begin{equation}
\Omega^{(1)} = \frac{1}{2} \sum_{\alpha\beta}
n_f(\epsilon_\alpha)n_f(\epsilon_\beta) \int d \bm{r_1}
 d \bm{r_2} U(\bm{r_1}-\bm{r_2}) \bigl
 |\mathcal{B}_{\alpha\beta}(\bm{r_1},\bm{r_2}) \bigr
 |^2 \,,
\label{OmegaC1}
\end{equation}
where $n_f(\epsilon)$ is the Fermi-Dirac distribution,
$\phi_\alpha(\bm{r})$ and $\epsilon_\alpha$ are exact
eigenfunctions and eigenenergies of the non-interacting
Hamiltonian~\eqref{e2}, and
\begin{equation}
\mathcal{B}_{\alpha\beta}(\bm{r_1},\bm{r_2}) =  \phi_\alpha(\bm{r_1})
\phi_\beta(\bm{r_2})
- \phi_\alpha(\bm{r_2}) \phi_\beta(\bm{r_1}) . \label{B1}
\end{equation}
After averaging over disorder, the contribution~\eqref{OmegaC1} becomes
\begin{equation}
\langle \Omega^{(1)} \rangle =
\int \frac{dE d\omega}{\Delta^2} n_f(E)n_f(E+\omega)  \int d \bm{r_1}
 d \bm{r_2} U(\bm{r_1}-\bm{r_2})
 \mathcal{K}_1(\bm{r_1},\bm{r_2},E,\omega) \,,
\label{OmegaC2}
\end{equation}
where $\Delta=1/\nu_d L^d$ stands for the mean-level spacing, $L$ is the
system size, and $\nu_d $ the density of states.
The function
\begin{equation}
\label{K1-definition}
\mathcal{K}_1
%(\bm{r_1},\bm{r_2},E,\omega)
= \frac{\Delta^2}{2}\sum_{\alpha\beta}\left \langle
\bigl |\mathcal{B}_{\alpha\beta}(\bm{r_1},\bm{r_2}) \bigr |^2
\delta(E+\omega-\epsilon_\alpha)\delta(E-\epsilon_\beta)\right \rangle
\end{equation}
describes correlations of two eigenstates of the non-interacting
Hamiltonian~\eqref{e2}.
In general, for the system size $L\to \infty$, the function
$\mathcal{K}_1$ exhibits the following scaling behavior
%(see Eq.(5.38) of Ref.~[\onlinecite{MirlinReview}])
\begin{equation}
\begin{split}
\mathcal{K}_1(\bm{r_1},\bm{r_2},E,\omega)  &= L^{-2d} \left
  (\frac{|\bm{r}_1-\bm{r}_2|}{L_\omega}\right)^{\mu_2}\tilde{\mathcal{K}}_1\left
  (\frac{|\bm{r}_1-\bm{r}_2|}{L_\omega}\right),
\\
\tilde{\mathcal{K}}_1(x) &= \begin{cases}
1,\qquad\qquad x\ll 1 , \\
x^{-\mu_2},\qquad  x\gg 1
\end{cases} \,,
\end{split} \label{K1C1}
\end{equation}
where $L_\omega = L (\Delta/|\omega|)^{1/d}$ is the length scale set
by the frequency $\omega$. For two adjacent in energy eigenstates one has
$\omega \sim \Delta$ and $L_\omega \sim L$, for larger frequencies the
scale  $L_\omega$ is much less than $L$.
It is important that the exponent $\mu_2$ is positive,
$\mu_2 > 0$, due to Hartree-Fock antisymmetrization of wave functions
in Eq.~\eqref{B1}. This means that for distances shorter than
$L_\omega$ the correlation function (\ref{K1C1}) is suppressed at
criticality as compared to the metallic system. This should be
contrasted to multifractal correlations of wave functions (without
antisymmetrization) that are enhanced at criticality. We will discuss
this point and, more generally, the spectrum of critical exponents
within the $\sigma$-model framework in Sec.~\ref{s2}.

We assume the electron-electron interaction of the following form:
\begin{equation}
\label{e4}
U(R) = u_0 \Bigl [ 1 +  (R/a)^2\Bigr ]^{-\lambda/2}\,,
\end{equation}
with $\lambda >d$.
This yields
%\begin{equation}
%\langle \beta\Omega^{(1)}\rangle =
%\frac{1}{\Delta^2} \int dE d\omega n_f(E)n_f(E+\omega)
%L^{-d} g \begin{cases}
%(a/L_\omega)^{d-\lambda+\mu_2} ,\qquad d+\mu_2 <\lambda \\
%1, \qquad d<\lambda \leqslant d+\mu_2 \\
%(L/L_\omega)^{d-\lambda},\qquad \lambda \leqslant d
%\end{cases}
%\end{equation}
\begin{equation}
\label{e5}
\langle \Omega^{(1)}\rangle  \propto \int \frac{dE d\omega}{\Delta}
n_f(E) n_f(E+\omega)  u(L_\omega) \,,
\end{equation}
where
\begin{equation}
u(L_\omega) = \nu_d u_0 a^d \begin{cases}
(a/L_\omega)^{\mu_2} ,&\qquad  d+\mu_2 <\lambda ,\\
(a/L_\omega)^{\mu_2} \ln \frac{L_\omega}{a}, & \qquad \lambda=d+\mu_2  ,\\
(a/L_\omega)^{\lambda-d}, & \qquad  d<\lambda < d+\mu_2 .
\end{cases}
\end{equation}
 %Here $S_d = ...$ stands for the area of the $d$-dimensional sphere.

Following Ref.~\cite{LeeWang}, we can consider $u(L_\omega)$ as
%%running in a sense of renormalization group dimensionless
renormalized interaction
parameter. It is seen that for $\lambda>d$ the
electron-electron interaction is indeed irrelevant near non-interacting fixed
point $u=0$.  While the above calculations are based on
evaluation of the Hartree-Fock contribution to the thermodynamic potential,
this conclusion is in agreement with the analysis of Ref.~\cite{PB}
based on Finkelstein non-linear $\sigma$-model.

\subsection{Dephasing rate: Second-order correction to the electron self-energy}

%%%%%%%%%%%%%%%%%%%%%%%%%%%%%%%%%%%%%%%%%%%%%%%%%%%%%%%%%%%%%%%%%%%%%%%%%%%%%
\begin{figure}[t]
\centerline{\includegraphics[width=120mm]{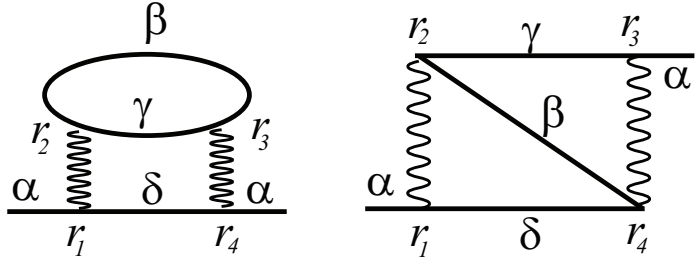}}
\caption{Second-order interaction contributions to the electron
  self-energy that determine the dephasing rate.
%The wavy line denotes the interaction and solid line stands
%  for the single particle Green function. %Left figure:
                                %$\Sigma_a$. Right figure: $\Sigma_b$
                                %.
} \label{Figure2}
\end{figure}
%%%%%%%%%%%%%%%%%%%%%%%%%%%%%%%%%%%%%%%%%%%%%%%%%%%%%%%%%%%%%%%%%%%%%%%%%%%%%

In order to compute the  dephasing rate (or, more accurately,
the out-scattering rate which in this case coincides with the
dephasing rate), one needs to evaluate the second-order interaction
correction to the imaginary part of the self-energy (see
Fig.~\ref{Figure2}). In spirit of  Ref.~\cite{AALR}, we define the
self-energy of a given single-particle state $\alpha$,
\begin{gather}
\Sigma_\alpha^R(\varepsilon) = \frac{1}{8}  \int d\bm{r_1}
d\bm{r_2}d\bm{r_3} d\bm{r_4}
U(\bm{r_1}-\bm{r_2})U(\bm{r_3}-\bm{r_4})
\mathcal{B}^*_{\alpha\beta}(\bm{r_1},\bm{r_2})
\mathcal{B}_{\delta\gamma}(\bm{r_1},\bm{r_2})
 \notag \\
\times  \mathcal{B}^*_{\gamma\delta}(\bm{r_3},\bm{r_4})
\mathcal{B}_{\beta\alpha}(\bm{r_3},\bm{r_4})  \sum_{\beta\gamma\delta}
\frac{n_f(\epsilon_\beta)[1-n_f(\epsilon_\gamma)]
+n_f(\epsilon_\delta)[n_f(\epsilon_\gamma)-n_f(\epsilon_\beta)]}{\varepsilon
  +\epsilon_\beta-\epsilon_\gamma-\epsilon_\delta+i0} .
\end{gather}
The next step is the evaluation of the imaginary part of the averaged
self-energy
\begin{equation}
\Sigma^R(E,\varepsilon) = \Delta \left \langle \sum_\alpha
  \Sigma_\alpha^R(\varepsilon)\delta(E-\epsilon_\alpha) \right \rangle .
\end{equation}
Since we consider the situation at criticality, where fluctuations may
be strong, the averaging is not always an innocent procedure. We will
return to the justification of the averaging in Sec.~\ref{s3.4}.

The result of averaging can be presented in the form
\begin{gather}
\Imag \Sigma^R(E,\varepsilon) = - \pi \left( \prod_{j=1}^4 \int d\bm{r_j}\right )
% d\bm{r_2}d\bm{r_3} d\bm{r_4}
U(\bm{r_1}-\bm{r_2})U(\bm{r_3}-\bm{r_4})
\int \frac{d\Omega d\varepsilon^\prime}{\Delta^3} \Bigl \{
n_f(\varepsilon^\prime+\Omega)  \notag \\
\times [1-n_f(\varepsilon^\prime)]+[n_f(\varepsilon^\prime)-
n_f(\varepsilon^\prime+\Omega)]n_f(\varepsilon+\Omega) \Bigr \}
\mathcal{K}_2(\{\bm{r}_j\},E,\varepsilon,\varepsilon^\prime,\Omega)\,, \notag \\
\label{SigmaEo1}
\end{gather}
where the correlation function
$\mathcal{K}_2(\{\bm{r}_j\},E,\varepsilon,\varepsilon^\prime,\Omega)$ is defined
as follows
\begin{gather}
\mathcal{K}_2(\{\bm{r}_j\},E,\varepsilon,\varepsilon^\prime,\Omega)   =
\frac{\Delta^4}{8} \Bigl \langle \sum_{\alpha\beta\gamma\delta}
\mathcal{B}^*_{\alpha\beta}(\bm{r_1},\bm{r_2})
\mathcal{B}_{\delta\gamma}(\bm{r_1},\bm{r_2})
 \mathcal{B}^*_{\gamma\delta}(\bm{r_3},\bm{r_4})
\mathcal{B}_{\beta\alpha}(\bm{r_3},\bm{r_4})  \notag \\
\times  \delta(E-\epsilon_\alpha)
\delta( \varepsilon^\prime+\Omega-\epsilon_\beta) \delta
(\varepsilon^\prime-\epsilon_\gamma )
\delta (\varepsilon+\Omega-\epsilon_\delta ) \Bigr \rangle .
\end{gather}

To determine the scaling of the dephasing rate, we now specify
characteristic values of energy variables. First of all, we are
interested in dephasing at the mass shell ($E=\varepsilon$) and at
characteristic energy $E\sim T$. Since we are not aiming at
calculating the numerical prefactor of order unity, we can simply set
$E=0$.  Second, we will see that the
characteristic values of the integral variables $\varepsilon'$ and $\Omega$
are set by the temperature.
Performing integration of the Fermi-function factor  over energy
$\varepsilon^\prime$, we find
\begin{gather}
\Imag \Sigma^R(0,0) \sim - \frac{1}{2\Delta^3} \left(
  \prod_{j=1}^4 \int d\bm{r_j}\right )
% d\bm{r_2}d\bm{r_3} d\bm{r_4}
U(\bm{r_1}-\bm{r_2})U(\bm{r_3}-\bm{r_4})
\int d\Omega\, \Omega\notag \\
\times \Bigl \{\coth \frac{\Omega}{2T} -\tanh
\frac{\Omega}{2T}\Bigr \}
\mathcal{K}_2(\{\bm{r}_j\},0,0,\varepsilon'\sim T,\Omega)  \,.
\label{SigmaEo2}
\end{gather}
%In order to estimate temperature dependence of the dephasing rate,
%$1/\tau_\varphi$, it is enough to consider the case $E=\varepsilon=0$:
%$1/\tau_\varphi=-\Imag  \Sigma^R(0,0)$.

In order to proceed further, we need to know the scaling behavior of
the function $\mathcal{K}_2$. Assuming that $\varepsilon'\sim\Omega$ (this
is sufficient for our purposes, as both these frequencies are of order
of temperature), we have for $| \bm{r}_1-\bm{r}_2|,
|\bm{r}_3-\bm{r}_4| \ll R \leqslant L_\Omega$ the scaling form \cite{LeeWang}
\begin{equation}
\mathcal{K}_2(\{\bm{r}_j\},0,0,\varepsilon'\sim\Omega,\Omega)  = L^{-4 d} \left
  (\frac{|\bm{r}_1-\bm{r}_2|}{R} \frac{|\bm{r}_3-\bm{r}_4|}{R}\right
)^{\mu_2} \left ( \frac{R}{L_\Omega}\right )^\alpha .
 \label{K2C1}
\end{equation}
where  $\bm{R}= (\bm{r}_1+\bm{r}_2-
\bm{r}_3-\bm{r}_4)/2$.
The scaling behavior~\eqref{K2C1} can be motivated as follows.
At distances $R\sim L_\Omega$ the correlations between wave functions
at $r_{1,2}$, on one hand, and $r_{3,4}$, on the other hand,
decouple. Thus, scaling of $\mathcal{K}_2$ reduces to that of a
product of two independent correlators $\mathcal{K}_1$, i.e.,
$(|\bm{r}_1-\bm{r}_2|/R)^{\mu_2}(|\bm{r}_3-\bm{r}_4|)/R)^{\mu_2}$.
The exponent $\alpha$ describes the scaling with respect to a
remaining scaling variable $R/L_\Omega$. For $R\gg L_\Omega$ the
correlations quickly (exponentially) decay, so that this range of $R$
is not important for the integral.

Combining Eqs.~(\ref{SigmaEo2}) and (\ref{K2C1}), we obtain
\begin{equation}
|\Imag \Sigma^R(0,0)| \sim \nu_d
T \int_0^T d\Omega \int_a^{L_\Omega} d\bm{R}\, u^2(R) \left (\frac{R}{L_\Omega}\right )^\alpha \,.
\label{SigmaEo3}
\end{equation}
The result of integration over ${\bf R}$ depends on the relations
between $\lambda$, $\mu_2$, $\alpha$ and $d$. 
Let us first assume that $\alpha>-d$. Then, if
$\lambda > d+\mu_2$, we find
\begin{equation}
\label{tau_phi_1}
\frac{1}{\tau_\varphi} \propto  \delta u^2(a)
\begin{cases}
(T/\delta)^{1+2\mu_2/d},\qquad\qquad d-2\mu_2+\alpha > 0 , \\
(T/\delta)^{1+2\mu_2/d} \ln (\delta/T), \quad d-2\mu_2+\alpha=0 , \\
(T/\delta)^{2+\alpha/d} , \qquad\qquad d-2\mu_2+\alpha <0 .
\end{cases}
\end{equation}
In the case $\lambda = d+\mu_2$, we obtain from Eq.~\eqref{SigmaEo3}
\begin{equation}
\label{tau_phi_2}
\frac{1}{\tau_\varphi} \propto  \delta u^2(a)
\begin{cases}
(T/\delta)^{1+2\mu_2/d} \ln^2(\delta/T),\quad \qquad d-2\mu_2+\alpha > 0 , \\
(T/\delta)^{1+2\mu_2/d} \ln^3 (\delta/T), \qquad d-2\mu_2+\alpha=0 , \\
(T/\delta)^{2+\alpha/d}, \quad\qquad d-2\mu_2+\alpha <0 .
\end{cases}
\end{equation}
Finally, if  $\lambda < d+\mu_2$, we get
\begin{equation}
\label{tau_phi_3}
\frac{1}{\tau_\varphi} \propto \delta u^2(a)
\begin{cases}
(T/\delta)^{-1+2\lambda/d},\qquad\qquad 3d-2\lambda+\alpha > 0 , \\
(T/\delta)^{-1+2\lambda/d} \ln (\delta/T), \quad 3d-2\lambda+\alpha=0 , \\
(T/\delta)^{2+\alpha/d} , \qquad \qquad 3d-2\lambda+\alpha < 0 .
\end{cases}
\end{equation}
Here $\delta =1/\nu_d a^d$ is the ultraviolet energy cutoff.

For $\alpha\leq -d$, the integration over $\Omega$ is dominated by
the infrared cutoff which should be chosen self-consistently at 
$1/\tau_\varphi$, yielding
 \begin{equation}
\label{tau_phi_4}
\frac{1}{\tau_\varphi} \propto 
\begin{cases}
T u^2(a) |\ln u(a)|,\qquad\qquad \alpha = -d , \\
\delta \left[T u^2(a)/\delta\right]^{-d/\alpha},  \qquad\qquad \alpha < -d .
\end{cases}
\end{equation}

Thus, the temperature behavior of the dephasing rate depends on the
exponents $\mu_2$ and $\alpha$ as well as on dimensionality $d$ and
the index $\lambda$ characterizing the decay of interaction.
The exponents  $\mu_2$ and $\alpha$ belong to a set of exponents
characterizing correlations of wave functions at criticality; their
values depend on the particular critical point under consideration.
In the next section we represent the correlation
function $\mathcal{K}_2$ in terms of the operators in
the non-linear sigma model. This will allow us to substantiate the
above scaling arguments by the RG analysis and to evaluate the
necessary exponents within the $2+\epsilon$ expansion.

\section{Field theoretical approach to wave function correlations}
\label{s2}

\subsection{Non-linear sigma model}
\label{s2.1}

The effective theory characterizing long-distance low-energy physics
of a disordered electronic system is the non-linear $\sigma$-model
\cite{wegner79}.
We will use its replica version; the same calculations can be
performed in the supersymmetric formulation.
The field variable of the theory is the matrix field $Q(\bm{r})$ of size
$2n\times 2n$ that obeys the non-linear constraint
$Q^{2}(\bm{r})=\bm{1}$. The non-linear sigma model action has the form
\begin{equation}
\label{Ssigma}
S_\sigma[Q] = -\frac{g}{8} \int d \bm{r} \tr(\nabla
Q)^{2} +
\frac{g h^2}{4} \int d
\bm{r} \tr Q \Lambda \,,
\end{equation}
where
\begin{equation}\label{Qrep}
\Lambda = \begin{pmatrix}
 1_{n} & 0 \\
  0 & - 1_{n}
\end{pmatrix} .
\end{equation}
The  parameter $g$ in front of the kinetic term
is identified with the dimensionless longitudinal conductance (in units of $e^2/h$). 
In the RG framework,
it is interpreted as a running coupling constant of the theory.
The second term in Eq.~(\ref{Ssigma}), which breaks the ${\rm U}(2n)$
symmetry down to ${\rm U}(n)\times {\rm U}(n)$, normally has an
imaginary prefactor proportional to frequency $\varepsilon$. We choose the
constant in front of this term to be real, which is more convenient
for RG analysis.

In the case of a 2D system in transverse magnetic field, there is an
additional contribution to the action that has a form of the
topological term ~\cite{PruiskenNLSM}
\begin{equation}
\label{S-topological}
S_{\rm top}[Q] = \frac{\theta}{8\pi} \int d \bm{r} \tr Q \nabla_x Q
\nabla_y Q\,.
\end{equation}
The angle $\theta$ in front
of the topological term is given by the
fractional part of the $2\pi g_{xy}$, where $g_{xy}$ stands
for the dimensionless Hall conductivity (in units $e^2/h$)~\cite{PruiskenBurmistrov}.

We now proceed by expressing the correlation function $\mathcal{K}_2$
in terms of eigenoperators of the $\sigma$-model. This calculation is
only based on the symmetry of the theory, so it is equally applicable
to the quantum Hall transition in 2D and to the Anderson transition in
$d>2$ dimensions.

\subsection{Expression for $\mathcal{K}_2$ in terms of the eigenoperators}
\label{s2.2}

It is convenient to express first the function $\mathcal{K}_2$ in
terms of the exact single-particle Green's function
$G_E^{R,A}(\bm{r},\bm{r}^\prime)$:
\begin{gather}
\mathcal{K}_2(\{\bm{r}_j\},E,\varepsilon,\varepsilon^\prime,\Omega)  =
\frac{\Delta^4}{\pi^4}
\Bigl \langle \Imag G_E^R(\bm{r}_4,\bm{r}_1) \Imag
G_{\varepsilon^\prime+\Omega}^R(\bm{r}_3,\bm{r}_2)\Bigl [
\Imag G_{\varepsilon+\Omega}^R(\bm{r}_1,\bm{r}_4) \notag \\
\times \Imag G_{\varepsilon^\prime}^R(\bm{r}_2,\bm{r}_3) -
\Imag G_{\varepsilon+\Omega}^R(\bm{r}_1,\bm{r}_3) \Imag
G_{\varepsilon^\prime}^R(\bm{r}_2,\bm{r}_4)\Bigr ]\Bigr \rangle .
\end{gather}
In order to extract the exponent $\alpha$, it is enough to consider
scaling of $\mathcal{K}_2(\{\bm{r}_j\},0,0,0,0)$ with the system size $L$ for
fixed $|\bm{r}_j-\bm{r}_k|$.
Then, following standard steps (see
e.g. Ref.~\cite{MirlinReview}), we obtain
\begin{gather}
\mathcal{K}_2(\{\bm{r}_j\},0,0,0,0)  = \frac{\Delta^4}{(2 \pi \gamma)^4}
\Biggl \langle \tr  \Bigl [ \Lambda Q_{ab}(\bm{r_1})\Lambda
Q_{ba}(\bm{r_4})\Bigr ]
\tr  \Bigl [ \Lambda Q_{cd}(\bm{r_2})\Lambda Q_{dc}(\bm{r_3})\Bigr ] \notag \\
+
\tr \Bigl [ \Lambda Q_{ab}(\bm{r_1})\Lambda Q_{bc}(\bm{r_4}) \Lambda
Q_{cd}(\bm{r_2})\Lambda Q_{da}(\bm{r_3})\Bigr ]
\Biggr \rangle \label{K2NLSM1}
\end{gather}
Here $\gamma=(\pi \nu_d)^{-1}$, the symbol $\tr$ stands for the trace over 
retarded-advanced space only and
$\langle \ldots \rangle$ denotes the averaging with the $\sigma$-model
action $S_\sigma$. It is important to stress that the replica indices $a, b,
c, d$ are fixed and different: $a\neq b \neq c \neq d$. It reflects
the fact that we are dealing with four different eigenstates.
When deriving Eq.~(\ref{K2NLSM1}), we have assumed that all distances
$|\bm{r}_j-\bm{r}_k|$ exceed the mean free path, so that the averaged
Green's functions between any two points can be neglected. (This
assumption does not affect the scaling but simplifies calculations.)
On the other hand, we can think about all points as located close to
each other from the point of view of the $\sigma$-model (say, on the
scale of several mean free paths), so that we omit space argument of $Q$
matrices in what follows. 

In view of the presence of the fixed replica indices, Eq.~\eqref{K2NLSM1}
is not convenient for further computations with the action $S_\sigma$
which is $U(n)\times U(n)$ invariant. In order to obtain a more
convenient, $U(n)\times U(n)$ invariant expression, we use the fact
that the correlation function is invariant with respect to global
unitary rotations in the replica space. This allows us to average
the expression in angular brackets in
Eq.~\eqref{K2NLSM1} over such global unitary rotations.
In general, the $U(n)\times U(n)$ invariant operators of the $k$th order in $Q$ matrices
have the form $O_{\lambda}=\Tr (\Lambda Q)^{k_1}\ldots \Tr (\Lambda Q)^{k_m},$
where $\lambda=\{k_1,\ldots,k_m\}$ is a partition of the number $k=k_1+\ldots+k_m$,
such that $k_1\geq k_2\ldots\geq k_m$. 
In particular, for $k=1$ we have one operator, $\{1\}$, for $k=2$ two operators
$\{2\}$ and $\{1,1\}$, for $k=3$ three operators $\{3\}$, $\{2,1\}$, 
and $\{1,1,1\}$, for $k=4$ five operators $\{4\}$, $\{3,1\}$, $\{2,2\}$, $\{2,1,1\}$, 
and $\{1,1,1,1\}$, and so on. 

Since the correlator $\mathcal{K}_2$, Eq.~(\ref{K2NLSM1}), contains four $Q$ matrices,
after the averaging over $U(n)\times U(n)$ global rotations 
the result is expressed in terms of the invariant operators of even orders $\leq 4$,
i.e. with $k=4$ and $k=2$ (and in addition a constant term corresponding to $k=0$).
Specifically, we obtain (see Appendix~\ref{Appendix3})
\begin{equation}
\mathcal{K}_2 = \frac{\Delta^4}{(2 \pi \gamma)^4} \left (
\sum_j C_j O_j[Q] +C_0 \right ) \,, \label{K2NLSM2}
\end{equation}
where operators $O_j[Q]$ and corresponding coefficients are given as
\begin{eqnarray}
O_4[Q] &=&  \Tr (\Lambda Q)^4, \label{refO4}\\
O_{3,1}[Q] &=&  \Tr (\Lambda Q)^3 \Tr \Lambda Q, \\
O_{2,2}[Q] &=& \Tr (\Lambda Q)^2\Tr (\Lambda Q)^2,\\
O_{2,1,1}[Q] &=&  \Tr (\Lambda Q)^2 (\Tr \Lambda Q)^2,\\
O_{1,1,1,1}[Q] &=& (\Tr \Lambda Q)^4, \\
O_{2}[Q] &=& \Tr (\Lambda Q)^2,\\
O_{1,1}[Q] &=& (\Tr\Lambda Q)^2  \label{refO11}
\end{eqnarray}
and
\begin{eqnarray}
C_4 &=& \frac{-3 + 13 n + 16 n^2 + 4 n^3}{4 n^2 (-1+n)(1 + n)^2 
(2+n)(3+n)}, \\
C_{3,1} &=& -\frac{-7 + 3 n + 12 n^2 + 4 n^3}{2 n^2 (-1 + n^2)^2 (3+n)(2+n)}, \\
C_{2,2} &=& \frac{-3 - 21 n + 20 n^2 + 32 n^3 + 8 n^4}{8 n^2 (-1 +
  n^2)^2 (3+n)(2+n)}, \\
C_{2,1,1} &=& -\frac{3 + 2 n}{2n^2 (-1 + n) (1 + n)^2 (3 + n)}, \\
C_{1,1,1,1} &=& \frac{5 + 5 n + 2 n^2}{8 n^2 (-1 + n^2)^2(3+n)(2+n)}, \\
C_2 &=& \frac{-3+3 n + 4n^2}{2n^2(-1 + n)^2 (3 +  n)}, \\
C_{1,1} &=&- \frac{3 + 5 n +4 n^2}{2n^2 (-1 + n^2)^2 (3+n)(2+n)}, \\
C_0 &=& \frac{-9+93n+4n^2-76n^3-24n^4}{2n(-1 + n^2)^2(3+n)(2+n)} .
\end{eqnarray}

While the operators $O_j[Q]$ are explicitly invariant with respect to
the ${\rm U}(n)\times {\rm U}(n)$ symmetry group of the
$\sigma$-model, they are {\it not} eigenoperators of
the renormalization group.  The eigenoperators that are linear combinations of
$O_j[Q]$ are in fact fully determined by the symmetry group (and realize its different
representations, similarly to conventional spheric function realizing
representations of the rotation group) \cite{Helgason}. 
The seven RG eigenoperators can be enumerated by Young frames:
 $P_{4}$, $P_{3,1}$, $P_{2,2}$, $P_{2,1,1}$, $P_{1,1,1,1}$, $P_{2}$, and $P_{1,1}$~\cite{wegner80,Wegner}.
To express these RG eigenoperators in terms of basis operators $O_j[Q]$  
one can use the results of Refs.~\cite{Wegner,Kravtsov}
 or the one-loop renormalization~\cite{wegner80,Pruisken85}. 
Finally, using Eq.~\eqref{K2NLSM2} we represent
the correlation function $\mathcal{K}_2$ as a linear combination of
the eigenoperators of RG,
\begin{gather}
\mathcal{K}_2 = \frac{\Delta^4}{(2 \pi \gamma)^4}
 \Biggl \langle \frac{4n^2}{4n^2(n^2-1)^2} P_{2,2}[Q]
 -\frac{4n^2+8n+3}{6n^2(n+1)^2(n^2+n-2)} P_{2,1,1}[Q]\notag
 \\+\frac{4n^2+16n+15}{2n^2(n+1)^2(n+2)(n+3)}P_{1,1,1,1}[Q]\Biggr \rangle .
 \label{KRep}
\end{gather}
It is remarkable that only three eigenoperators (out of seven)
enter the obtained expression for the correlation function
$\mathcal{K}_2$. The eigenoperators $P_{4}$, $P_{3,1}$ and $P_{2}$ with leading singularities in the replica limit $n=0$
do not contribute to  $\mathcal{K}_2$.

\section{Temperature dependence of dephasing rate at criticality}
\label{s3}

\subsection{Anderson transition in $2+\epsilon$ dimensions}
\label{s3.1}

In $d=2+\epsilon$ dimensions with small $\epsilon$ the Anderson
transition takes place at
large dimensionless conductance $g$. This allows one to
explore the critical behavior within the $\epsilon$-expansion.
The $\beta$-function governing the renormalization of the conductance
is known up to the five-loop order \cite{Hikami,WegnerB}
\begin{equation}
\label{e3.1.1}
-\frac{d t}{d\ln L} = \beta(t) =  \epsilon t - 2 t^3 -6t^5  + O(t^7) \,,
\end{equation}
where $t= 1/2\pi g$.
The metal-insulator transition occurs at the critical point $t_\star$
defined by $\beta(t_\star)=0$ that yields
\begin{equation}
\label{e3.1.2}
t_* = \left(\frac{\epsilon}{2}\right)^{1/2} - \frac{3}{2}
\left(\frac{\epsilon}{2}\right)^{3/2} + O (\epsilon^{5/2})\,.
\end{equation}
The localization length index $\nu$ reads
\begin{equation}
\label{e3.1.3}
\nu = - 1/\beta'(t_*) = \frac{1}{2\epsilon} - \frac{3}{4} +
O(\epsilon)\,.
\end{equation}

Further, according to Ref.~\cite{Wegner}, the anomalous dimensions of the
eigenoperators $P_j$ are given up to four-loop order as
\begin{equation}
\label{e3.1.4}
\gamma_{P_j}(t) = a_2 \rho(t) +\zeta(3) c_3 t^4 + O(t^5)
,\qquad \rho(t) = t + \frac{3}{2}(n^2+1)t^3+\frac{n}{3}(n^2+7)t^4
\end{equation}
where coefficients $a_2$ and $c_3$ are summarized in
Table~\ref{Table1}. It is worthwhile to mention that the coefficient
$c_3$ is proportional to $a_2$. Therefore, in the replica limit
$n\to 0$,  the anomalous dimensions of the eigenoperators involved in
Eq.~\eqref{KRep} become
\begin{eqnarray}
\gamma_{P_{2,2}}(t) &=& O(t^5) ,\\
\gamma_{P_{2,1,1}}(t) &=& -4t -6t^3+24 \zeta(3)t^4 +O(t^5) , \label{g1}\\
\gamma_{P_{1,1,1,1}}(t) &=& -12 t -28 t^3+216 \zeta(3)t^4 +O(t^5) . \label{g2}
\end{eqnarray}

The exponent $\mu_2$ is determined by the anomalous dimension of the
eigenoperator $P_{1,1}$ \cite{wegner80,Pruisken85}, see
Appendix~\ref{Appendix2}. In the replica limit $n\to 0$,
\begin{equation}
\label{e3.1.5}
\gamma_{P_{1,1}}(t) = -2 t -3 t^3+6 \zeta(3)t^4 +O(t^5) .
\end{equation}
Using Eq.~\eqref{e3.1.2}, we find the exponent $\mu_2$ in $2+\epsilon$
dimensions~\cite{Pruisken85,Wegner},
\begin{equation}
\mu_2 = -\gamma_{P_{1,1}}(t_\star)  = \sqrt{2\epsilon} - \frac{3}{2}
\zeta(3)\epsilon^2 + O(\epsilon^{5/2}) .
\end{equation}
The exponent $\alpha$ is determined by the maximal value of the
anomalous dimensions (with oppposite sign) of operators $P_{2,2}$,
$P_{2,1,1}$ and $P_{1,1,1,1}$ at the critical point:
\begin{equation}
\alpha = \max \{-\gamma_{P_{2,2}}(t_\star),
-\gamma_{P_{2,1,1}}(t_\star),-\gamma_{P_{1,1,1,1}}(t_\star)\}.
\end{equation}
The anomalous dimensions of the operators $P_{2,1,1}$ and $P_{1,1,1,1}$
are negative, whereas $\gamma_{P_{2,2}}$ vanishes
within the known accuracy:
\begin{equation}
\alpha = O(\epsilon^{5/2}) .
\end{equation}

%%%%%%%%%%%%%%%%%%%%%%%%%%%%%%%%
\begin{table}[t]
\begin{center}
\caption{Coefficients of $\gamma$-functions from Ref.~\cite{Wegner}.}
\begin{tabular}{c|c|c}
& $a_2$&$c_3$\\
\hline
$P_{4}$&$12-8n$&$12(3-2n)(2-n)(3-n)$\\
$P_{3,1}$&$4-8n$&$4(1-2n)(6-13n+n^2)$\\
$P_{2,2}$ &$-8n$& $84 n^2$ \\
$P_{2,1,1}$&$-4-8n$&$4(1+2n)(6+13n+n^2)$\\
$P_{1,1,1,1}$&$-12-8n$&$12(3+2n)(2+n)(3+n)$\\
\hline
$P_{2}$&$2-4n$&$2 (1 - 2 n) (1 - n) (3 - n)$\\
$P_{1,1}$&$-2-4n$&$2 (1 + 2 n) (1 + n) (3+ n)$\\
\end{tabular}
\label{Table1}
\end{center}
\vspace{1cm}
\end{table}
%%%%%%%%%%%%%%%%%%%%%%%%

Therefore, in the case $\lambda>d+\sqrt{2\epsilon}$ (for brevity we
keep here only the leading, one-loop, contribution to $\mu_2$),
the temperature behavior of the dephasing rate is given by
\begin{equation}
\frac{1}{\tau_\phi} \propto \delta u^2(a)  \left
  (\frac{T}{\delta}\right )^{1+\sqrt{2\epsilon}} .
\label{Eq2E}
\end{equation}
In the special case $\lambda= d+\sqrt{2\epsilon}$,
\begin{equation}
\frac{1}{\tau_\varphi} \propto  \delta u^2(a)  \left
  (\frac{T}{\delta}\right )^{1+\sqrt{2\epsilon}} \ln^2(\delta/T).
\end{equation}
Only in the regime $d<\lambda<d+\sqrt{2\epsilon}$, the temperature
behavior of the dephasing rate is determined by the exponent
$\lambda$:
\begin{equation}
\frac{1}{\tau_\varphi} \propto  \delta u^2(a)  \left
  (\frac{T}{\delta}\right )^{-1+2\lambda/d} .
\end{equation}

Let us focus on the case of the ``most short-range'' interaction,
$\lambda > d + \mu_2$, when $\tau_\phi^{-1} \propto T^p$ with $p = 1+
2\mu_2/d$. The scaling of the dephasing length is then given by 
\begin{equation}
\label{L_phi}
L_\phi \propto T^{-1/z_T}\ ; \qquad\qquad z_T= \frac{d}{p} = \frac{d}{1+2\mu_2/d}\,.
\end{equation}
The exponent $z_T$ belongs to a class of dynamical critical exponents,
as it governs the scaling of a characteristic length scale (dephasing
length) with a variable having the dimension of energy. It is
worth emphasizing that the localization probems possess rich
physics and are in general characterized by several dynamical
exponents. This fact has not always been appreciated in the
literature. In the present case, one should distinguish the exponent
$z_T$ controlling the scaling with temperature from the exponent
$z$ governing the scaling with frequency. The latter exponent
has a trivial (non-interacting) value $z = d$ for the
short-range interaction.

The transition width induced by inelastic scattering scales as
$T^\kappa$, where the exponent $\kappa$ is found by comparison of the
dephasing length (\ref{L_phi}) with the localization (correlation) length $\xi
\propto |P-P_c|^{-\nu}$, where $P$ is the parameter, driving the
transition (e.g., electron concentration or disorder strength). This
yields
\begin{equation}
\label{kappa}
\kappa = \frac{1}{z_T \nu} = \frac{1 + 2\mu_2/d}{\nu d}\,.
\end{equation}
Substituting the above formulas for $\mu_2$ and $\nu$, we get the
following results for the dynamical exponent $z_T$  and for the index
$\kappa$ up to four-loop order of the
$\epsilon$-expansion:
\begin{eqnarray}
\label{zT-epsilon}
z_T  &=& 2 - 2\sqrt{2}\epsilon^{1/2} + 5\epsilon  -
4\sqrt{2}\epsilon^{3/2} + O(\epsilon^2)    \,; \\[0.3cm]
\label{kappa-epsilon}
\kappa &=& \epsilon + \sqrt{2}\epsilon^{3/2} + \epsilon^2 +
\epsilon^{5/2}/\sqrt{2} + O(\epsilon^3)    \,.
\end{eqnarray}

\subsection{Integer quantum Hall transition}
\label{s3.3}

We have employed numerical
techniques in order to calculate ${\mathcal K}_{1,2}$ and the
corresponding exponents for the quantum Hall transition.
Our computations are based on the
Chalker-Coddington network. We extract eigenvalues and eigenfunctions
near zero (pseudo)energy from large square systems. In the present context
the system size $L$ is parametrized with the number of links
$N={512,768,1024}$ in each direction.
Typically of the order of $10^4$ wavefunctions enter the ensemble
average for the correlators ${\mathcal K}_{1,2}$.

\begin{figure}[t]
\centerline{\includegraphics[width=100mm]{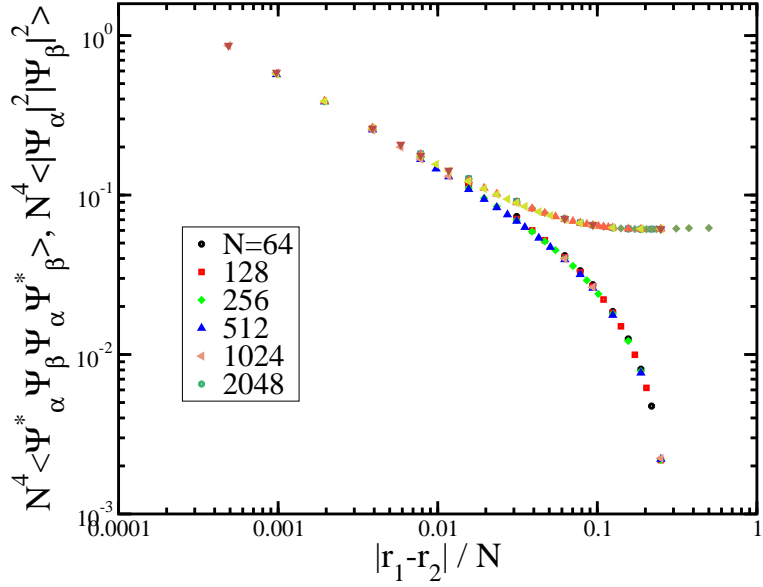}}
\caption{ Hartree and Fock contributions to the wave function correlator
  ${\mathcal K}_1$ as defined in Eq. (\ref{K1-definition}) for pairs
  of wavefunctions neighboring in energy. It is seen
  that in the scaling regime both correlation functions
follow the same power law, including the exponent and the amplitude.
The corresponding exponent $\Delta_2 \simeq - 0.52$
is known from the multifractal analysis \cite{evers08}.}
\label{f3}
\end{figure}

In Fig.~\ref{f3} we show separately two correlation functions (Hartree
and Fock terms), the difference of which constitutes  the wave
function correlator ${\mathcal K}_1$ defined in
Eq. (\ref{K1-definition}). It is nicely seen that in the scaling
regime of point separation $|{\bf r_1}-{\bf r_2}|$ much smaller than
the system size $N$ both correlation functions follow the same power
law. The corresponding exponent is well known from the multifractal
analysis of moments of wave functions, $\Delta_2 \simeq - 0.52$.
It is important that not only the exponent but also the prefactor
is the same. This ensures that the difference between the two
correlation functions scales with another (subleading) exponent
$\mu_2$.

%%%%%%%%%%%%%%%%%%%%%%%%%%%%%%%%%%%%%%%%%%%%%%%%%%%%%%%%%%%
%%%%%%%%%%%%%%%%%%%%%%%%%%%%%%%%%%%%%%%%%%%%%%%%%%%%%%%%%%%%%%%
%%%%%%%%%%%%%%%%%%%%%%%%%%%%%%%%%%%%%%%%%%%%%%%%%%%%%%%%%%%
\begin{figure}[t]
\centerline{\includegraphics[width=100mm]{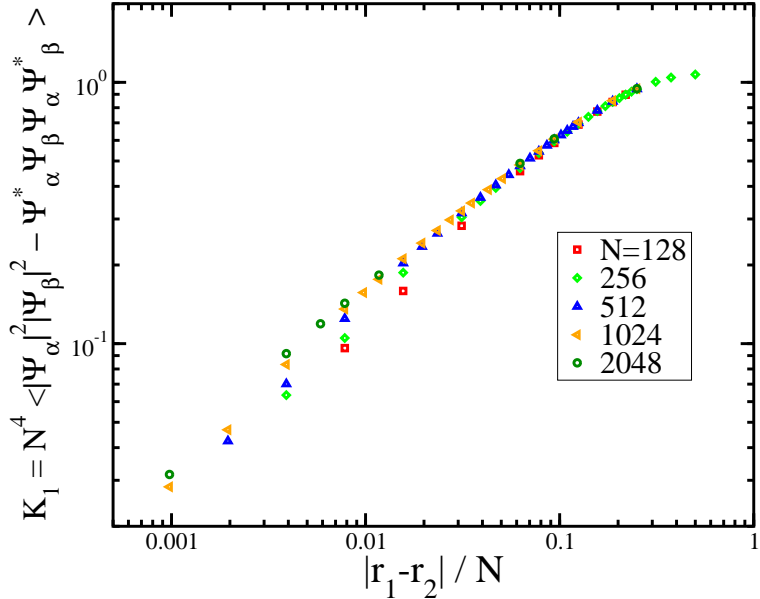}}
\caption{Correlator ${\mathcal K}_{1}$ representing the difference of the
two functions shown in Fig. \ref{f3}. The leading
power laws cancel and ${\mathcal K}_{1}$ is determined by the subleading
contributions. Two types of deviations from pure power-law behavior
(straight line in the double-log scale) are seen.
At small distances the data collapse is not perfect due
to corrections to scaling originating from the ultraviolet cutoff scale
(lattice constant). At large distances deviations from the power-law
scaling are caused by $|{\bf r_1}-{\bf r_2}|$ approaching $N$.
The extracted value of the power-law exponent is $\mu_2\simeq 0.62\pm
0.05$.} \label{f4}
\end{figure}
%%%%%%%%%%%%%%%%%%%%%%%%%%%%%%%%%%%%%%%%%%%%%%%%%%%%%%%%%%%

Figure \ref{f4} shows the
correlator ${\mathcal K}_{1}$ representing the difference of the
two functions shown in the previous plot, Fig. \ref{f3}. As emphasized
above, the leading power-law contributions to the Hartree and Fock
terms cancel, so that ${\mathcal K}_{1}$ is determined by subleading
contributions. For a pure power-law scaling behavior we would have a
straight line in the double logarithmic scale. We observe two types of
deviations from this behavior. At large distances ($|{\bf r}_1 - {\bf r}_2|/N \gtrsim 0.1$)
there is a considerable curvature
which is related to higher subleading terms.  At small distances the
data collapse is not perfect due to deviations from scaling related to
the ultraviolet cutoff scale (lattice constant) $a$ which leads to
emergence of an additional scaling parameter $a/|{\bf r_1}-{\bf
  r_2}|$. The numerical analysis yields the power-law exponent
$\mu_2\simeq 0.62\pm 0.05$.

\begin{figure}[t]
\centerline{\includegraphics[width=100mm]{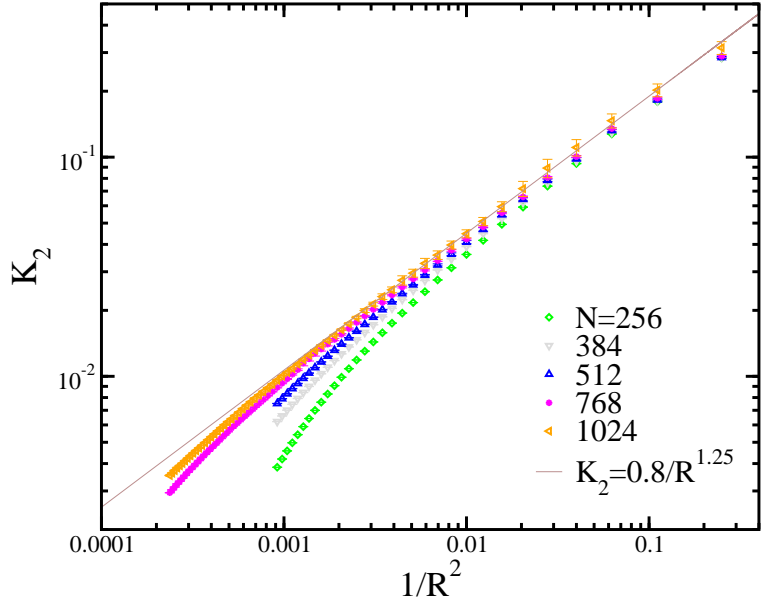}}
\caption{Correlator ${\mathcal K}_{2}$ at fixed small distances
$|{\bf r}_1-{\bf r}_2|=|{\bf r}_3 - {\bf r}_4|$ as a function of
  the distance $R$ between the pairs ${\bf r}_{1,2}$ and ${\bf
    r}_{3,4}$. Evolution of  ${\mathcal K}_{2}$ with system
  size $N$ is shown for a quartett of neighboring energies.
  The average is over $10^6$ samples with two quartetts selected
  for each sample; the error bars give one standard deviation
  to indicate the residual statistical uncertainty.
  It is seen that with increasing $N$ the data
  approach the straight line, corresponding to a power-law dependence.
  The corresponding exponent is $\simeq 1.25$, i.e. equal to $2\mu_2$
  within the numerical uncertainty,
  implying that $\alpha \simeq 0$.  } \label{f5}
\end{figure}

In Fig.~\ref{f5} we show the correlation function
${\mathcal K}_{2}$ at fixed small distance between pairs of the points,
$|{\bf r}_1-{\bf r}_2|=|{\bf r}_3 - {\bf r}_4| \equiv \rho$, as a
function of the
distance $R$ between the pairs. When the system size $N$ increases,
the data approach a straight line, corresponding to a power-law
dependence on $R$ with the exponent $\simeq 1.25$. This exponent is
equal to $2\mu_2$ within the uncertainty of our numerical analysis. Therefore,
within our accuracy the exponent $\alpha$ defined in Eq. (\ref{K2C1})
is indistinguishable from zero.

%%%%%%%%%%%%%%%%%%%%%%%%%%%%%%%%%%%%%%%%%%%%%%%%%%%%%%%%%%%
%%%%%%%%%%%%%%%%%%%%%%%%%%%%%%%%%%%%%%%%%%%%%%%%%%%%%%%%%%%

%%%%%%%%%%%%%%%%%%%%%%%%%%%%%%%%%%%%%%%%%%%%%%%%%%%%%%%%%%%
\begin{figure}[t]
\vskip0.3cm
\centerline{\includegraphics[width=100mm]{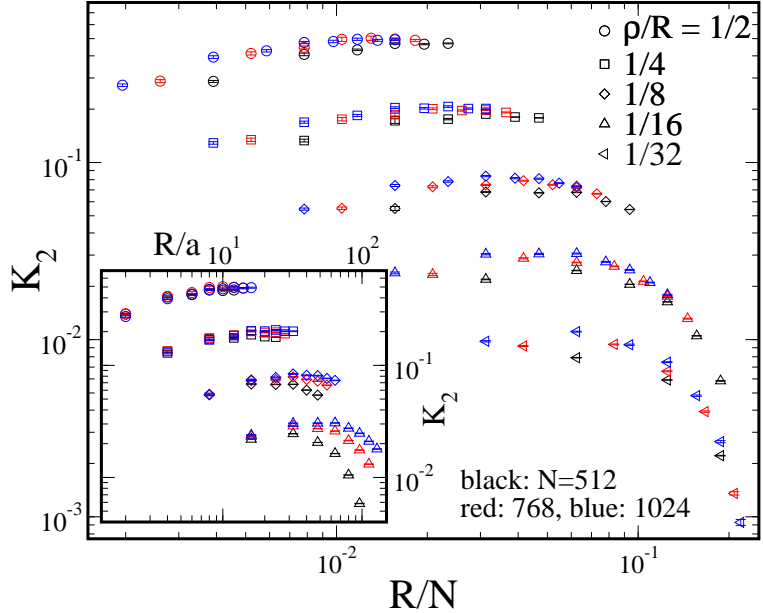}}
\caption{ Scaling behavior of ${\mathcal K}_{2}$ as a function of
  $R/N$ for a fixed distance ratio, $\rho/R = 1/2, 1/4, 1/8, 1/16,
  1/32$. Different colors correspond to different system sizes
(black: 512, red: 768, blue: 1024).
  According to Eq.~(\ref{K2C1}), we expect the scaling $\propto
  (R/N)^\alpha$ for $R/N \ll 1$. This scaling behavior is represented
  by a plateau at intermediate values of $R/N$, yielding $\alpha$
  close to zero.
At larger $R/N$ we see deviations
  from a straight line, since other contributions (scaling with higher
  irrelevant  exponents) become significant. At small $R$ deviations
  from the data collapse are due to corrections in $a/\rho$. {\it
    Inset:} same data replotted as a function of $R/a$. At small
  distances no $N$-dependence is observed. This confirms that
  deviations from a simple power-law scaling (straight line) are
  controlled by $a/\rho$.  }
\label{f6}
\end{figure}
%%%%%%%%%%%%%%%%%%%%%%%%%%%%%%%%%%%%%%%%%%%%%%%%%%%%%%%%%%%

The above result on the exponent $\alpha$ is further supported by
Fig.~\ref{f6} where the correlation function ${\mathcal K}_{2}$ is
plotted as a function of $R/N$ for fixed $\rho/R$. According to
Eq. (\ref{K2C1}), this is a direct way to determine the exponent
$\alpha$. We see that at $R/N \lesssim 0.1$ the plots are almost flat,
which implies $\alpha \simeq 0$.
At small $R$ we observe again the
deviations from scaling controlled by the parameter $a/\rho$, as
demonstrated in the inset.

\begin{figure}[t]
\centerline{\includegraphics[width=135mm]{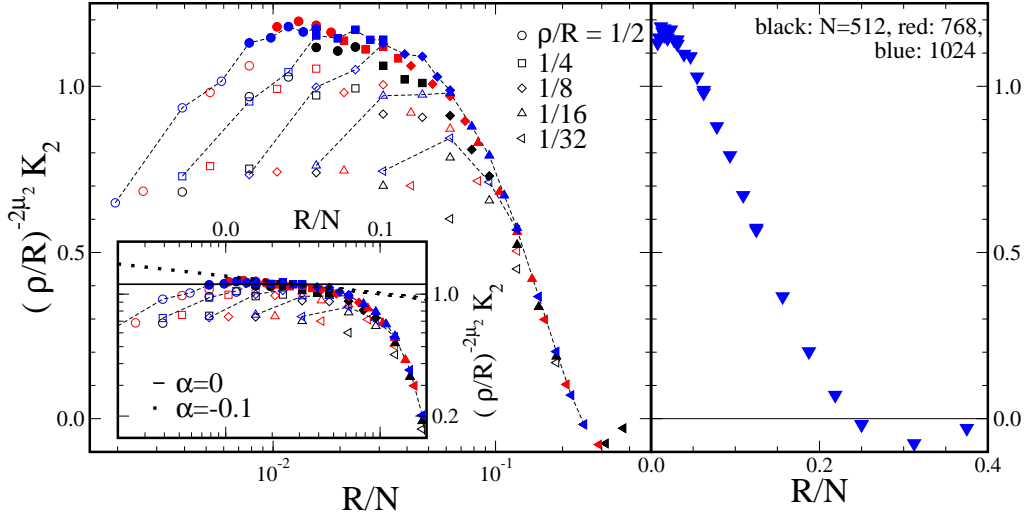}}
\caption{\emph{Left panel:} Scaling of $(\rho/R)^{-2\mu_2}{\mathcal K}_2$ assuming a critical
  index $\mu_2=0.625$. Different symbols correspond to different
  values of the ratio $\rho/R$; different colors to different system
  sizes $N$, as in Fig.~\ref{f6}.
Deviations from the single parameter scaling for smallest
  $\rho$ (open symbols) are due to corrections controlled by
  $a/\rho$. {\it Inset:} Same data on a log-log scale. Solid line (corresponding to $\alpha=0$)
and dotted line (corresponding
  to $\alpha=-0.1$) indicate the power-law scaling region and
show the uncertainty range for the numerical value of the exponent $\alpha$.
\emph{Right panel:} Single-parameter scaling function (as
  obtained from the data points shown by full symbols in the left
  panel) on the double-linear scale.}
 \label{f7}
\end{figure}

To make the scaling properties of ${\mathcal K}_{2}$ particularly
clear, we replot in the left panel of Fig.~\ref{f7} the data of Fig.~\ref{f6},
multiplying each trace by $(\rho/R)^{-2\mu_2}$ with $2\mu_2 =
1.25$. It is seen that all data (full symbols) collapse on a single-parameter
 scaling curve showing the dependence of ${\mathcal
  K}_{2}(\rho/R)^{-2\mu_2}$ on $R/N$. This confirms that the value of
$\mu_2$ is correct. The open symbols show the data
points for three smallest values of $\rho$; they deviate from the
scaling curve due to corrections in $a/\rho$. The plateau of the
single-parameter scaling curve at small values of $R/N$ 
%corresponds to $\alpha \simeq 0$ 
yields $\alpha \simeq -0.05\pm 0.1$ (see the inset in the left panel of Fig.~\ref{f7} 
where the same data are shown on a log-log scale).
\footnote{Extracting the bounds on
possible values for $\alpha$ from our numerical data is complicated by the fact
%The reason is
that, within the system sizes available to us, corrections in $a/\rho$
and $R/L$ are not negligible,
see inset of Fig. \ref{f6} and
the discussion at the end of this section \ref{s3.3}.
%The scaling analysis Fig. \ref{f7} 
%does not really help in this respect,
%because there an 
The uncertainty in numerical determination of
$\mu_2$ enters the scaling analysis (Fig. \ref{f7}) as an additional
source of uncertainty in $\alpha$.
%A conservative estimate based on the plateau curvature
%in the data sets, Fig. \ref{f6},
%with $\rho/R=1/2,1/4,1/8$ at largest
%system size $N=1024$ would be $\alpha\approx -0.05\pm0.1$ which is consistent
%with the analysis in the inset of Fig. \ref{f7}.
} The single-parameter
scaling curve is also shown in the right panel of Fig.~\ref{f7} on the double-linear scale.

%%%%%%%%%%%%%%%%%%%%%%%%%%%%%%%%%%%%%%%%%%%%%%%%%%%%%%%%%%%
%%%%%%%%%%%%%%%%%%%%%%%%%%%%%%%%%%%%%%%%%%%%%%%%%%%%%%%%%%%

To summarize, we have found the values $\mu_2  \simeq 0.62 \pm 0.05$
and $\alpha \simeq -0.05\pm 0.1$ for the critical exponents describing the
correlation functions of interest. These values nicely agree with those
previously obtained by Lee and Wang \cite{LeeWang}. It should be
stressed, however, that we used systems of a linear size an order of
magnitude larger than in Ref.~\cite{LeeWang}. The large system sizes $N
\gtrsim 500$ were crucially important for our scaling analysis shown
in Fig.~\ref{f7}. Indeed, it is seen there that the scaling function
reaches the expected power-law behavior $\propto (R/N)^\alpha$ (with
$\alpha$ close to zero) only at relatively small values of the scaling
argument,  $R/N \lesssim 0.05$. On the other hand, the scale $\rho$
cannot be too small, since the corrections controlled by the parameter
$a/\rho$ become substantial unless $\rho/a \gtrsim 4$. Finally, we have
a condition $R \gg \rho$, which in practice requires that $R/\rho
\gtrsim 2$. Combining all this, we see that in order to have a window
of good power-law scaling, we need $N$ considerably larger than $2
\times 4 \times 20 = 160$. Our system sizes reasonably satisfy this
requirement. As a result, we get a window of $R/N$ approximately
between 0.007 and 0.05, where the required scaling takes place.
On the other hand, relatively small system sizes did not allow the
authors of Ref.~\cite{LeeWang} to obtain this scaling window. Indeed,
their Fig.~2 shows a function that changes by several orders of
magnitude without any developed saturation plateau. While the authors
of  Ref.~\cite{LeeWang} argued correctly in favor of small $\alpha$,
it seems difficult to make a reliable conclusion concerning $\alpha$
with only data for small systems (as shown in their Fig.~2) at hand.

Using the obtained values of $\mu_2$ and $\alpha$, we can calculate
the exponents controlling the interaction effects, see
Eqs.~(\ref{L_phi}) and (\ref{kappa}). Assuming the case of ``most
short-range interaction'', $\lambda > 2 + \mu_2 \simeq 2.62$, we get
\begin{equation}
\label{qhe-dyn-exp}
p \simeq 1.62\ ; \qquad z_T \simeq 1.23\ ; \qquad \kappa \simeq 0.346
\,.
\end{equation}
While calculating $\kappa$ in Eq.~(\ref{qhe-dyn-exp}), we used the
value of the localization length exponent, $\nu \simeq 2.35$ found by
Huckestein and coauthors (see the review \cite{huckestein95})
and confirmed by several later works. Recently, Slevin and Ohtsuki
\cite{slevin09}
reconsidered the problem and concluded that corrections to scaling are
much larger that was previously thought. As a result, they obtained a larger
value of the localization length critical exponent, $\nu \simeq
2.59$. If this value as used, we get a somewhat smaller result for the
exponent $\kappa$,
\begin{equation}
\label{qhe-kappa-modified}
\kappa \simeq 0.314
\,.
\end{equation}
We will compare these values to existing experimental results in
Sec.~\ref{s4}, where we will also discuss expected modifications in the case
of Coulomb interaction.

\subsection{Anderson transition in $d=3$ and higher dimensions}
\label{s3.2}

For the Anderson transition in $d=3$ the
$\epsilon$-expansion cannot give quantitatively reliable predictions
for critical exponents. Therefore, one has to rely on numerical
simulations. The critical exponent of the localization length for the
unitary symmetry class was found to be $\nu = 1.43 \pm 0.04$
\cite{slevin97}. While the most relevant multifractal exponents
(characterizing the scaling of moments of wave function amplitudes)
$\Delta_q$ have been extensively studied, no numerical analysis of the
subleading exponents, in partiuclar $\mu_2$ and $\alpha$, has been done
to the best of  our knowledge. This is an interesting direction for
future research. In particular, an intriguing question is whether the
condition $d-2\mu_2+\alpha>0$ becomes violated in 3D (or, more generally
at sufficiently high dimensionality). According to
Eq.~(\ref{tau_phi_1}), this would result in a change of the behavior
of the dephasing rate $1/\tau_\phi$ that would become dependent on
the exponent $\alpha$.

\subsection{Discussion: Consistency of the  calculation of the
  dephasing rate and the transition width}
\label{s3.4}

Having completed the calculation of the dephasing rate and of the
localization transition width, we have to come back to the assumptions
made in course of the calculation and check their
consistency. Specifically, we have made two important assumptions: on
the averaging procedure and on the critical character of the wave
functions involved.

\subsubsection{Critical character of wave functions}
\label{s3.4.1}

When we calculated the dephasing rate as the imaginary part of the
self energy ${\rm Im}\: \Sigma(E,\varepsilon)$, we assumed that the zero
energy (i.e. the position of the chemical potential $\mu$) is exactly
at the critical point. On the other hand, the obtained transition
width scales as $\propto T^\kappa$ with $\kappa < 1$, i.e. it is much
larger than $T$. Therefore, it is important to check that finite
deviation of $\mu$ from criticality,
$|\mu-\mu_c| \propto T^\kappa$ does not invalidate the calculation. This
detuning from criticality will produce a characteristic scale
(localization or correlation length), $\xi_\mu \propto
|\mu-\mu_c|^{-\nu}$ that determines the range of critical behavior of
correlation functions. Inserting here $|\mu-\mu_c| \propto T^\kappa$,
we get $\xi_\mu \propto T^{-1/z_T}$, which is just the condition
$\xi_\mu = L_\phi$ that we used to determine the transition width.
Therefore, for characteristic detuning from critical energy $\sim
|\mu-\mu_c|$ the critical correlations extend up to $L_\phi\propto T^{-1/z_T}$.
On the other hand, the range of spatial
integration in Eq.~(\ref{SigmaEo3}) was set by the thermal length
$L_T\propto T^{-1/d}$. We thus have to compare $L_\phi$ and
$L_T$. Since $z_T < d$, we have $L_\phi \gg L_T$, which justifies the
calculation.

\subsubsection{Averaging of dephasing rate}
\label{s3.4.2}

While evaluating $\tau_\phi$, we have performed averaging over the
disorder realizations. A natural question to ask is whether this is a
valid procedure. Indeed, we know that in the localized phase such an
averaging breaks down with lowering temperature \cite{fleishman80}
since the level spacing of those states with which the given state is
coupled becomes larger than the averaged dephasing
rate. As a consequence the Golden-rule calculation of the dephasing
rate that treats these states essentially as a continuum breaks down.
However, in the critical regime we are considering the situation is
essentially different. As we discussed in Sec.~\ref{s3.4.1}, the
localization length $\xi_\mu$, which is equal to the dephasing
$L_\phi$, increases (with lowering temperature) faster than
$T^{-1/d}$. As a result, the number of states contributing to
the dephasing rate of a given state---i.e. of states located  in the
energy interval of width $T$ and in the spatial volume (area for $d=2$)
of extension $\xi_\mu = L_\phi\propto T^{-1/z_T}$
increases as a power law with decreasing
$T$. In this situation, the dephasing rate $\tau_\phi$ is a
self-averaging quantity.

\subsubsection{Other possible contributions to dephasing}
\label{s3.4.3}

Strictly speaking, what we have calculated in this paper is the 
lowest-order (golden rule) dephasing rate governed by the generic range 
of frequencies ($\Omega\sim \varepsilon' \sim  T$).  Thus, putting it 
rigoristically, we have calculated the upper bound on $L_\phi$, and, 
consequently, the lower bound on $z_T$ and upper bound on $\kappa$.  Indeed, 
formally one cannot exclude the possibility that a contribution of a 
different scaling domain of frequency and spatial variables to the 
lowest-order dephasing rate or contribution of 
higher order in $U$ is larger. Our preliminary analysis indicate that this 
does not happen (apart from possible logarithmic corrections to scaling) 
for critical points with relatively small anomalous 
exponents, like Anderson transition in $2+\epsilon$ dimensions and 
quantum Hall transition which are in the focus of this paper. It is a 
challenging task to understand whether such exotic contributions may 
dominate the dephasing rate at transitions with strong fluctuations, 
like Anderson transition in higher dimensionalities. The complexity of 
the problem is related to the fact that the other contributions are 
controlled by wave function correlations characterized by  
different scaling exponents.  We postpone this analysis to future work.

\section{Coulomb interaction and experiment}
\label{s4}

Most of this paper is devoted to the case of short-range interaction,
when the interaction irrelevant in the RG sense and
the dephasing rate is controlled by critical properties at the
non-interacting fixed point.
In this section we briefly discuss the present understanding of the
case of long-range Coulomb interaction and summarize the
available experimental results.

\subsection{Criticality at Anderson and quantum Hall transitions with
  Coulomb interaction}
\label{s4.1}

For long-range ($1/r$) Coulomb interaction, the dephasing rate is
proportional to temperature,  $1/\tau_\phi \sim T$. It is easy to see
how the short-range interaction results develop into this behavior
with decreasing $\lambda$. Indeed, the lowest possible ``short-range''
$\lambda$ is $\lambda =d$, for which the first line of
Eq.~(\ref{tau_phi_3}) should be used, yielding $1/\tau_\phi \sim
T$. Further decreasing $\lambda$ does not change the scaling of the
dephasing rate anymore, since $1/\tau_\phi$ cannot vanish slower than
temperature in a system with meaningfully defined fermionic
excitations. The result $1/\tau_\phi \sim T$ implies that the scaling
with temperature is the same as with frequency, which is usually the case
at ``standard'' quantum phase transitions. Therefore, in contrast to
the case of short-range interaction, for the
long-range interaction there is no need in distinguishing the
dynamical exponents governing the frequency and temperature scaling, $z_T=z$.

On the other hand, the long-range interaction problem is characterized
by several dynamical exponents controlling the frequency scaling of
different observables \cite{finkelstein90,belitz94}.
The reason for this complex behavior is the existence
of several conserved quantities. Let us assume for simplicity that the
system is spin-polarized (or else, the spin invariance is completely
broken, e.g., by spin-orbit interaction or by magnetic
impurities). Then the conserved quantities are particle number and
energy. As a consequence, there are two Goldstone modes, which are
characterized by poles of the diffusion type but with a non-trivial
scaling, $q^{z_i} \sim \omega$. In notations of Ref.~\cite{belitz94}
$z_1$ corresponds to the energy mode, and $z_3$ to the density mode.

The exponent $z_1$, which was denoted as $\zeta$ by Finkelstein
\cite{finkelstein90} and as $2+\gamma^*$ in Ref.~\cite{baranov02},
governs the renormalization of frequency in the
$\sigma$-model action. It is this exponent that plays a role of
$z_T=z$ for the problem we are considering, i.e., it controls the
scaling of the dephasing length $L_\phi \sim \tau_\phi^{1/z}$ and
therefore enters the formula
\begin{equation}
\label{kappa-coulomb}
\kappa = 1/\nu z \ ; \qquad \qquad z \equiv z_1
\end{equation}
for the exponent $\kappa$ of the transition width.

We emphasize the distinction between different dynamical exponents, since
this was not always appreciated by researchers in the field and has
led to confusions and controversies. A number of authors used the
exponent $z_3$, which is equal to 1 for the case of $1/r$ interaction,
instead of $z_1$ in Eq.~(\ref{kappa-coulomb}). This is incorrect. To
make this point clear, it is constructive to draw an analogy with a 2D
problem: the critical problem in
$2+\epsilon$ dimensions with small $\epsilon$ bears a lot of
similarities with a 2D problem with large conductance. The exponent
$z_3 =1$ exists also in 2D: it controls the plasmon pole in the
(reducible) density response function. It is well known, however, that
the plasmon pole does not affect the conductivity; in
particular, the dephasing rate is recalculated in dephasing length
according to a diffusion formula $L_\phi \propto (D\tau_\phi)^{1/2}$,
corresponding to $z=2$. Furthermore, the localization effects are
controlled by cooperons and by ``delayed diffusons''
\cite{polyakov98,ludwig08} that are given by
ladder diagramms without interaction vertex corrections. These are
just modes that acquire the $z_1$ dynamical scaling at criticality.

In $2+\epsilon$ dimensions for the considered symmetry class (time
reversal and spin symmetries are broken; denoted as ``MI(LR)'' in
Ref.~\cite{belitz94} ), the $\beta$-function, the critical point, and
the  exponents $\nu$ and $z$ are known up to the
two-loop order \cite{baranov02}:
\begin{eqnarray}
\label{beta-coulomb}
\beta(t)  & = & \epsilon t  - 2 t^2 - 4 A t^3
\ ; \qquad\qquad A\simeq 1.64\ ;
\\[0.2cm]
\label{t-coulomb}
t_* & = & \frac{\epsilon}{2} - \frac{A}{2}\epsilon^2 + O(\epsilon^3)
\\[0.2cm]
\label{nu-coulomb}
\nu & = & \frac{1}{\epsilon} - A + O(\epsilon) \ ;\\[0.2cm]
z & = & 2 + \frac{\epsilon}{2} +
\left(\frac{A}{2} - \frac{\pi^2}{24} - \frac{3}{4} \right) \epsilon^2
+ O(\epsilon^3) \,.
\label{z-coulomb}
\end{eqnarray}
Substituting (\ref{nu-coulomb}) and   (\ref{z-coulomb}) into
(\ref{kappa-coulomb}), we get
\begin{equation}
\label{kappa-coulomb-epsilon}
\kappa =  \frac{\epsilon}{2} + \left(\frac{A}{2} - \frac{1}{8}\right)\epsilon^2 
+ O(\epsilon^3) \,.
\end{equation}

For the quantum Hall transition with Coulomb interaction not much can
be said concerning the values of the exponents on the theoretical
level. Since the transition happens in the strong-coupling regime of
the $\sigma$-model, no reliable analytical predictions can be
made. The numerical analysis is also very difficult: (i) exact
diagonalization can be performed for small systems only, which is not
suffucient for determining the critical behavior, and (ii) no
approximate method that would allow to get the interacting critical
exponents has been developed. On general grounds, and using the
analogy with the Anderson transition in $2+\epsilon$ dimensions, one
can say that there is no reasons to expect that the exponents $\nu$
and $z$ (as well as any other exponents) would take the same values as
for the non-interacting system. Also, there is no reasons for the
exponent $z \equiv z_1$ to take any ``simple'' value (like, e.g., 1 or
2).

\subsection{Experiments on localization transition in interacting systems}
\label{s4.2}

\subsubsection{Anderson transition in 3D}
\label{s4.2.1}

The 3D localization transition was extensively studied on doped semiconductor
systems, such as Si:P, Si:B, Si:As, Ge:Sb. In most of the works, samples with a
substantial degree of compensation [i.e. acceptors in addition to donors,
e.g. Si:(P,B)] were used, which allows one to vary the amount of disorder and
the electron concentration independently. On these samples,
values of the conductivity exponent $s$ in the vicinity
of $s\approx 1$ were reported
\cite{thomas82}
%%zabrodskii84,field85,hirsch88}
with scattering of values and
the uncertainties of the order of $10\%$. A similar result was obtained
for an amorphous material Nb$_x$Si$_{1-x}$. We recall that $s$ is
expected to be equal in 3D to the localization length exponent $\nu$
according to the scaling relation $s=\nu(d-2)$.

On the other hand, the early study of the transition in undoped Si:P
\cite{paalanen82}
%%,thomas83,rosenbaum83}
gave an essentially different result,
$s\approx 0.5$, which is also in conflict with the Harris inequality
$\nu> 2/d$. This discrepancy was resolved in
\cite{stupp93} where it was found that the actual
critical region in an uncompensated Si:P is rather narrow and that the scaling
analysis restricted to this range yields $s\approx 1.3$. A more recent study
along these lines \cite{waffenschmidt99}
%, which employed uniaxial stress to tune through the transition (as used in
%\textcite{paalanen82,thomas83,rosenbaum83}),
%has essentially confirmed these conclusions,
yielded  $s=1.0\pm 0.1$, in agreement with the values obtained for
samples with compensation. It thus appears that for the orthogonal symmetry class
(preserved time-reversal and spin invariances) the experiments have
converged to the value $s=1.0\pm 0.1$.  (The only exception is a recent
experiment on uncompensated Si:B \cite{bogdanovich99} where a larger
value was found, $s\approx 1.6$. A possible explanation is that the
temperatures reached in this work were not sufficiently low. Another
possibility is that Si:B belongs to a different universality class, in
view of stronger spin-orbit scattering.) Results on the dynamical
scaling are much scarcer: it was found to be $z=2.94\pm 0.3$ in
Ref.~\cite{waffenschmidt99} and $z\approx 2$ in Ref.~\cite{bogdanovich99}.

The fact that the experimental value $s\approx 1$ differs from the
result $s=\nu\simeq 1.57\pm 0.02$ \cite{slevin99} for non-interacting systems of the
orthogonal symmetry class
is in line with the general expectation that the Coulomb interaction
affects the critical exponents.

\subsubsection{Quantum Hall transition}
\label{s4.2.2}

Experiments on the integer quantum Hall plateau transition
%yield the following results for the critical
%exponents. First, the index $\nu$ of the localization length is found
%to be $\nu=2.3{\pm} 0.1$ \cite{koch91}; 
determine 
%this value was confirmed more
%recently in Ref.~\cite{hohls01}. Second, 
the width of the critical
region (peak in $\sigma_{xx}$ and plateau transition in $\sigma_{xy}$), which
scales with the temperature $T$ as $\Delta B {\propto} T^\kappa$.
The early measurements of the exponent $\kappa$ performed on InGaAs/InP samples 
yielded $\kappa {=} 0.42{\pm} 0.04$ \cite{wei88}.
A number of works discussed the effect of macroscopic inhomogeneities
\cite{van-schaijk00,karmakar04,li05} that complicate observation of the true IQH critical
behavior.  The final conclusion is  \cite{li05} that for short-range disorder,
when the true IQH criticality can be achieved, $\kappa {=} 0.42{\pm}
0.01$, in agreement with the result of Ref.~\cite{wei88}.
A very recent work \cite{li09} where the quantum Hall transition in AlGaAs/AlGaAs samples
was
analyzed down to very low temperatures ($1\:{\rm mK}$) confirmed this
result for $\kappa$. 

Experimental determination of the exponents $\nu$ and $z_T$ is a highly 
complicated problem. The values reported in the literature are   $\nu \simeq 
2.3$  \cite{koch91,hohls01} and $z_T\simeq 1$  \cite{li09}, with the 
scattering of data of the order of  $10\%$.  Our feeling, however, is that 
these data might be essentially affected by systematic errors.  Specifically, 
the works  \cite{koch91,hohls01} where $\nu$ was measured  reported the 
values of $\kappa$ in the range from 0.6 to 0.8 (i.e. much larger than the 
true exponent 0.42).  It is understood now that such increased values of 
$\kappa$ correspond to situations where macroscopic inhomogeneities do not 
allow one to observe the true quantum Hall criticality. In such a situation 
the observed $\nu$ may also differ from the actual quantum Hall value. 
Further, the most direct determination of $z$  \cite{li09} was based on the 
analysis of the data for different system sizes, which is a nice way to find 
the dephasing length. However,  there may be a problem \cite{comment} related to the fact 
that disorder strength was apparently correlated with the sample width (see 
Fig.3a of Ref.~\cite{li09}). As a result, the natural temperature scale (mean 
free time) for different samples is different. This is not a small effect: as 
is seen from Fig.3a of Ref.~\cite{li09} changing the sample width by factor 
of 5 not only changes the saturation temperature by factor of  $\approx 5$ 
but simultaneously changes the characteristic temperature scale by factor of 
$\approx 2.5$. In our view, this might considerably affect the determination 
of the dynamical exponent $z_T$. It seems that more experimental work may be 
needed to overcome these difficulties related to systematic errors in 
evaluation of  $z_T$ and $\nu$.

%Also, this work analyzes the scaling with sample
%size and comes to the conclusion that $z\simeq 1$. The authors do not
%give the error bars but the inspection of their data suggests that
%the accuracy of the last result is of order $10\%$ (assuming that larger
%systematic errors are excluded).

Summarizing the experimental findings, the most updated values of the
exponents are  $\nu=2.3{\pm} 0.1$, $\kappa {=} 0.42{\pm} 0.01$, and
$z_T \equiv z \simeq 1.0 \pm 0.1$, although the error in determination 
of $\nu$ and $z_T$ may be considerably underestimated due to systematic errors.
Let us remind the reader that the
theoretical results for the case of short-range interaction are as
follows: the value of $\nu$ ranges from 2.35 to 2.59, $z_T \simeq
1.23$, and $\kappa = 1/\nu z_T$ is in the range from 0.314 to
0.346. 
It appears that the difference in values of the exponents
between the cases of short-range (theory) and long-range (experiment) 
interactions is not so
large: $\lesssim 10\%$ for $\nu$, $\lesssim 20\%$ for $z_T$, and
$\lesssim 30\%$   for $\kappa$. 
(Again, for $\nu$ and $z_T$ might be large due to systematic errors.)
Nevertheless, the difference demonstrates
that the current experiments on criticality at quantum Hall transitions
cannot be explained in terms of the non-interacting fixed point.
An experimental realization of the
short-range interaction universality class remains a challenging issue
for future research.

%Finally, the frequency scaling of the transition width was
%found to be $\Delta B {\sim} \omega^\zeta$, with $\zeta=0.41{\pm} 0.04$
%\cite{engel93}. A more recent work \cite{hohls02} yields a result
%consistent with this value, but with somewhat larger uncertainty,
%$\zeta {=} 0.5{\pm} 0.1$.

\section{Conclusions}

To summarize, we have studied the scaling properties of dephasing rate at
critical point of the localization transition.
We have considered the case of a short-range interaction in systems
with no spin degeneracy (or broken spin-rotation symmetry). In this
situation the interaction is found to be RG-irrelevant, and the
critical properties can be studied by performing the scaling analysis
near the non-interacting fixing point.

More specifically, we  considered problems with broken time-reversal
invariance: the quantum Hall transition and the Anderson transition in
$2+\epsilon$  dimensions. Our work combined analytical and numerical
analysis. In the analytical part, we used the framework of the
non-linear $\sigma$ model. We identified operators controlling
the scaling of the correlation function that determines the dephasing
rate. Further, we performed their RG analysis in $2+\epsilon$
dimensions. This allowed us to find the analytical results for the
exponents $p$, $z_T$, and $\kappa$ governing the temperature scaling
of the dephasing rate, dephasing length, and the transition width.

The numerical analysis was used to obtain critical exponent at the
quantum Hall transition.  Our results for the exponents largely agree
with those obtained in Refs.~\cite{LeeWang,Wang2}. However, our system
sizes are much largely than those studied in Ref.~\cite{LeeWang} that
was crucial for getting a window of distances where corrections to
power-law scaling are small.

Experimental results on localization transition with short-range
interaction are extremely desirable. In the case of quantum Hall
transition one could imagine screening of the Coulomb interaction by an
external gate. It would be extremely interesting to observe the change
of the exponents compared to the case of long-range interaction. This
is a challenging task, especially since
the difference between the exponents appears to be not so large  (see
Sec.~\ref{s4.2.2}).
For 3D Anderson transition, its experimental realization and investigation
in cold-atoms systems would be of great importance.

On the theoretical side,  the wave function correlation
exponents $\mu_2$ and $\alpha$ need to be
evaluated for 3D Anderson transition (with and without time-reversal
invariance), in order to predict the behavior of the dephasing rate
and thus the exponents  $p$, $z_T$, and $\kappa$.
More generally, investigation of the statistics of wave functions at
criticality beyond the leading multifractal behavior represents an
important research field. It would be interesting to understand the
evolution of this statistics (which includes
the statistics of Hartree-Fock matrix elements) from the regime of weak to strong
multifractality, as it has been done for the spectrum of leading
multifractal exponents $\Delta_q$ \cite{evers08}.

\section{Acknowledgements}

The work was supported by RFBR Grant Nos. 09-02-12206 and
09-02-00247, the Council for grants
of the Russian President Grant No. MK-125.2009.2, RAS Programs
``Quantum Physics of Condensed Matter'',
``Fundamentals of nanotechnology and nanomaterials'', the Russian
Ministry of Education and Science under contract No. P926, 
by the Center for Functional Nanostructures of the Deutsche
Forschungsgemeinschaft, by the SPP ``Graphene'' of the DFG, 
and by the EUROHORCS/ESF EURYI Awards scheme.
%{\bf To check and complete:  (i) financial agencies, (ii) discussions
%  with ...}  
S.B. and F.E. thank I. Kondov for support in optimizing the computer code
used for the numerical simulations.  I.S.B. is grateful to A. Pruisken for useful discussions.   
I.S.B. is
grateful to the Institute of Condensed Matter Theory and Institute of
Nanotechnology of Karlsruhe Institute of Technology for hospitality.

\appendix
%%%%%%%%%%%%%%%%%%%%%%%%%%%%%%%%%%%%%%%%%%%%%%%%%%%%%%%%%%%%%%%%%%%%%%
%         APPENDIX 1
%
%%%%%%%%%%%%%%%%%%%%%%%%%%%%%%%%%%%%%%%%%%%%%%%%%%%%%%%%%%%%%%%%%%%%
%\section{\label{Appendix1} Appendix. Comparison with Finkelstein
%  non-linear sigma model}

%%%%%%%%%%%%%%%%%%%%%%%%%%%%%%%%%%%%%%%%%%%%%%%%%%%%%%%%%%%%%%%%%%%%%%
%         APPENDIX 2
%
%%%%%%%%%%%%%%%%%%%%%%%%%%%%%%%%%%%%%%%%%%%%%%%%%%%%%%%%%%%%%%%%%%%%
\section{\label{Appendix2} Correlation function $\mathcal{K}_1$.}

In this Appendix we demonstrate that the function $\mathcal{K}_1$
corresponds to the eigenoperator $P_{1,1}$.
The function $\mathcal{K}_1$
can be expressed in terms of the exact single-particle Green's
functions $G_{R,A}(\bm{r},\bm{r}^\prime)$:
\begin{equation}
\mathcal{K}_1=\frac{\Delta^2}{\pi^2} \Biggl \langle \Imag G_{E+\omega}^R(\bm{r}_1,\bm{r}_1) \Imag G_{E}^R(\bm{r}_2,\bm{r}_2) - \Imag G_{E+\omega}^R(\bm{r}_2,\bm{r}_1)
\Imag G_{E}^R(\bm{r}_1,\bm{r}_2)\Biggr \rangle .
\end{equation}
Following standard steps (see e.g. Ref.~\cite{MirlinReview}), we obtain
\begin{equation}
\mathcal{K}_1= \frac{\Delta^2}{(2\pi\gamma)^2} \left  \langle \tr \Lambda Q_{aa}(\bm{r}_1)\tr\Lambda Q_{bb}(\bm{r}_2) +\tr \Lambda Q_{ab}(\bm{r }_1)\Lambda
Q_{ba}(\bm{r}_2)  \right \rangle \,,
\end{equation}
where repica indices $a, b$ are different, $a\neq b$, which reflects the
fact that we are dealing with two different eigenstates. Since we are
interested in the case for which two points are close to each other
(say, $|\bm{r}_1-\bm{r}_2|$ is of order of a few lattice constants
$a$), we can consider the arguments of $Q$ matrices as equal.
Below the arguments are omitted.

Let us define two operators bilinear in $Q$:
\begin{equation}
\tilde{O}_{\pm}[Q] =
 \tr  \Lambda Q_{ab} \Lambda Q_{ba} \pm \tr \Lambda Q_{ab} \tr \Lambda Q_{ba} .
 \label{Opm1}
\end{equation}
Clearly, $\mathcal{K}_1 \propto \tilde{O}_+$.
In order to obtain the $U(n)\times U(n)$ invariant expression let us
perform the following global rotation in the $Q$-matrix space,
\begin{equation}
Q(\bm{r}) \to U^{-1} Q(\bm{r}) U, \qquad U^{pp^\prime}_{ab} = U^p_{ab}
\delta^{pp^\prime} , \label{GT}
\end{equation}
which does not change the action $S_\sigma$. Thus, we introduce the
averaged, operators
\begin{equation}
O_{\pm}[Q] =
\left \langle \tr  \Lambda [U^{-1} Q U]_{ab} \Lambda [U^{-1} Q U]_{ba}
  \pm \tr \Lambda [U^{-1} Q U]_{ab} \tr \Lambda [U^{-1} Q U]_{ba}
\right \rangle_U   \label{Opm2}
\end{equation}
where $\langle \dots \rangle_U$ denotes averaging over $U(n)\times
U(n)$ global rotations. Since the action $S_\sigma$ is invariant under $U(n)\times
U(n)$, the $\sigma$-model averages of $O_{\pm}[Q]$ and
$\tilde{O}_{\pm}[Q]$ are equal,
\begin{equation}
\langle \tilde{O}_{\pm}[Q] \rangle = \langle O_{\pm}[Q]\rangle ,
\end{equation}

In order to perform the averaging over $U(n)\times U(n)$ rotations, we
use the following results~\cite{Mello}:
\begin{gather}
\langle (U^{-1})^{p}_{a\alpha} U^p_{\beta b}\rangle_U =V_1
\delta_{ab}\delta_{\alpha\beta} , \label{V1} \\
\langle (U^{-1})^{p}_{a\alpha} U^p_{\beta b} (U^{-1})^{p}_{c\gamma}
U^p_{\mu d} \rangle_U =
V_{1,1} \left [ \delta_{ab}\delta_{\alpha\beta}\delta_{cd} \delta_{\gamma\mu} +
 \delta_{ad}\delta_{\alpha\mu}\delta_{bc} \delta_{\beta\gamma}\right ] \notag \\
  +V_2 \left [ \delta_{bc}\delta_{\alpha\beta}\delta_{da} \delta_{\gamma\mu} +
 \delta_{ab}\delta_{\alpha\mu}\delta_{cd} \delta_{\beta\gamma}\right ]
\,,
\label{V2}
\end{gather}
where $V_1$, $V_{1,1}$ and $V_2$ are given in Table~\ref{Table2}.
The index $p = \pm$ distinguishes between the retarded and advanced
sectors, i.e. $(U^+,U^-)$ is an element of $U(n)\times U(n)$. The
averaging over  $U^+$ and $U^-$ is carried out independently.

Performing the averaging, we obtain the following $U(n)\times U(n)$
invariant results:
\begin{gather}
O_{\pm}[Q] = \frac{2n\pm 1}{2n^2(n\pm 1)} \left \{ \Tr
  (\Lambda Q)^2 \pm (\Tr\Lambda Q)^2 \right \} \pm \frac{(2n-1)\mp 2n}{2n^2(n\pm
  1)} \Tr \mathbf{1} .
\end{gather}
It is not difficult to check that operators $O_\pm$ are eigenoperators
under the action of the renormalization group: $O_-=P_{2}$ and $O_+ = P_{1,1}$ .

%%%%%%%%%%%%%%%%%%%%%%%%%%%%%%%%%%%%%
%%%%%%%%%%%%
\begin{table}[t]
\begin{center}
\begin{tabular}{c}
$V_1 = \frac{1}{n}$, \\
$V_{1,1} = \frac{n}{n (n^2 - 1)}$, \qquad $V_2 = -\frac{1}{n (n^2 - 1)}$,  \\
$V_{1,1,1} = \frac{n^2 - 2}{n (n^2 - 1) (n^2 - 4)}$,\qquad $V_{2,1} = -\frac{n}{n (n^2 - 1) (n^2 - 4)}$, \qquad
$V_3 = \frac{2}{n (n^2 - 1) (n^2 - 4)}$, \notag \\
$V_{1,1,1,1}= \frac{6 - 8 n^2 + n^4}{n^2 (n^2 - 1) (n^2 - 4) (n^2 - 9)}$, \,
$V_{2,1,1} = \frac{n (4 - n^2)}{n^2 (n^2 - 1) (n^2 - 4) (n^2 - 9)}$,\,
$V_{2,2} = \frac{6 + n^2}{n^2 (n^2 - 1) (n^2 - 4) (n^2 - 9)}$,\notag \\
$V_{3,1} = \frac{2 n^2 - 3}{n^2 (n^2 - 1) (n^2 - 4) (n^2 - 9)}$, \,
$V_4= -\frac{5 n}{n^2 (n^2 - 1) (n^2 - 4) (n^2 - 9)}$
\end{tabular}
\caption{Coefficients $V_j$ for averaging over the unitary group $U(n)$.}
\label{Table2}
\end{center}
\vspace{1cm}
\end{table}
%%%%%%%%%%%%%%%%%%%%%%%%%%%%%%%%%%%%%%%

%%%%%%%%%%%%%%%%%%%%%%%%%%%%%%%%%%%%%%%%%%%%%%%%%%%%%%%%%%%%%%%%%%%%%%
%         APPENDIX 3
%
%%%%%%%%%%%%%%%%%%%%%%%%%%%%%%%%%%%%%%%%%%%%%%%%%%%%%%%%%%%%%%%%%%%%
\section{\label{Appendix3}  Correlation function $\mathcal{K}_2$.}

In this Appendix, we express the correlation function $\mathcal{K}_2$
in terms of basis operators $O_j[Q]$. This requires, in
addition to bilinear operators considered in Appendix~\ref{Appendix2},
introducing operators of the fourth order in $Q$.
As in the previous Appendix, it is convenient to perform global
transformation~\eqref{GT} and average over $U(n)\times U(n)$
rotations. Then $\mathcal{K}_2$ becomes
\begin{equation}
\mathcal{K}_2 = \frac{\Delta^4}{(2\pi\gamma)^4} \Bigl \langle R_1[Q]
+R_2[Q]\Bigr \rangle
\end{equation}
where
 \begin{gather}
 R_1[Q] = \langle \tr   \Lambda (U^{-1} Q U)_{ab} \Lambda (U^{-1} Q U)_{ba}
\tr  \Lambda (U^{-1} Q U)_{cd}\Lambda (U^{-1} Q U)_{dc} \rangle_U ,
 \\
R_2[Q]= \langle\tr \Lambda  (U^{-1} Q U)_{ab}\Lambda  (U^{-1} Q
U)_{bc} \Lambda  (U^{-1} Q U)_{cd}\Lambda  (U^{-1} Q U)_{da} \rangle_U
..
\end{gather}
In order to perform averaging over $U(n)\times U(n)$ rotations it is
convenient to use diagrammatic technique developed in
Ref.~\cite{Brouwer}. The result is as follows
\begin{eqnarray}
R_1[Q] &=& \sum_{p=\pm} \Bigl \{
V_{2,2} (\Tr Q_p)^4 +  (4V_{4}+2V_{2,1,1}) \Tr Q_p^2 (\Tr Q_p)^2\notag \\ && +8
V_{3,1} \Tr Q_p^3 \Tr Q_p +(2V_{2,2}+V_{1,1,1,1}) \Tr
Q_p^2\Tr Q_p^2  \notag \\&&+ (4V_{2,1,1}+2V_{4})\Tr Q_p^4- 4 V_1 V_{2,1} (\Tr
Q_p)^2 \Tr A_p  \notag \\&&- 4 V_1 V_{1,1,1} \Tr Q_p^2 \Tr A_p
-8 V_1 V_{2,1} \Tr(A_p Q_p^2)  \notag \\&&- 8 V_1 V_{3} \Tr Q_p \Tr A_p Q_p 
+2 (V^2_{1,1}+ V^2_2) (\Tr A_p)^2 \notag \\&&+4 V_{1,1} V_2 \Tr A_p^2  
+ V_{1,1}^2 \Tr Q_p^2 \Tr Q^2_{-p}\notag \\&&
+ 2 V_2 V_{1,1} \Tr Q^2_p (\Tr Q_{-p})^2 +V_{2}^2 (\Tr Q_p)^2 (\Tr Q_{-p})^2
\Bigr \} ,
\end{eqnarray}
and
\begin{eqnarray}
R_2[Q]&=& \sum_{p=\pm} \Bigl \{
V_4 (\Tr Q_p)^4 +  (4V_{3,1}+2V_{2,2}) \Tr Q_p^2 (\Tr
Q_p)^2 \notag \\
&&+4(V_4+V_{2,1,1}) \Tr Q_p^3 \Tr Q_p +(V_4+2V_{2,1,1}) \Tr Q_p^2\Tr Q_p^2 
\notag \\ &&+
(V_{2,2}+V_{1,1,1,1}+4V_{3,1})\Tr Q_p^4- 4 V_1 V_3 (\Tr Q_p)^2 \Tr A_p
\notag \\ &&- 4 V_1 V_{2,1} \Tr Q_p^2 \Tr A_p- 4 V_1(V_3+V_{1,1,1}) \Tr(A_p Q_p^2) \notag \\ &&- 8 V_1 V_{2,1} \Tr Q_p \Tr A_p
Q_p+ 2 V_2^2 \Tr Q_p \Tr Q_{-p} \Tr A_p
\notag \\&& + 4 V_2 V_{1,1} \Tr Q_p \Tr B_p +2 V_{1,1}^2 \Tr B_p Q_p  +2 V_{1,1}
V_2 (\Tr A_p)^2 \notag \\ &&
+(V_{1,1}^2+V_2^2) \Tr A_p^2
\Bigr \} \,,
\end{eqnarray}
where the coefficents $V_{j_1,\ldots,j_m}$ are introduced in analogy
with Eqs.~(\ref{V1}), (\ref{V2}). Specifically, $V_{j_1,\ldots,j_m}$
arises when one averages a product of $j = j_1+ \ldots + j_m$ matrix elements
of $U$ and $j$ matrix elements of $U^{-1}$ as a coefficients in front
of terms corresponding to $m$ ``cycles'' of the sizes $j_1, \ldots j_m$.
Further,  we have introduced the notations
\begin{eqnarray}
Q_p &=& \frac{1+p\Lambda}{2} Q
\frac{1+p\Lambda}{2} , \\
A_p &=& \frac{1+p\Lambda}{2} Q   \frac{1-p\Lambda}{2}Q
\frac{1+p\Lambda}{2} ,\\
B_p &=& \frac{1+p\Lambda}{2} Q   \frac{1-p\Lambda}{2}Q
\frac{1-p\Lambda}{2}Q   \frac{1+p\Lambda}{2}  .
\end{eqnarray}
Using the expressions for the coefficents $V_{j}$ from Table~\ref{Table2}, we
can express the operators $R_1$ and $R_2$ via operators
\eqref{refO4}-\eqref{refO11}. The result is given in Eq.~\eqref{K2NLSM2}.


\begin{thebibliography}{100}

\bibitem{evers08} F.~Evers and A.D.~Mirlin, Rev. Mod. Phys. {\bf 80},
  1355 (2008); A. D. Mirlin, F. Evers, I. V. Gornyi and
  P. M. Ostrovsky, in {\it 50
  years of Anderson localization},
  ed. by E. Abrahams (World Scientific, 2010), p. 107;
  Int. J. Mod. Phys. B {\bf 24},  1577 (2010).

\bibitem{AA85} B.L.~Altshuler and A.G.~Aronov, in {\it
    Electron-Electron Interactions in Disordered Systems}, edited by
  A.L. Efros and M. Pollak (Elsevier, 1985), p.1.

\bibitem{finkelstein90} A.M.~Finkelstein, Sov. Sci. Rev. Sect. A {\bf
    14}, 1 (1990); in {\it 50 years of Anderson localization},
  ed. by E. Abrahams (World Scientific, 2010), p. 385;
  Int. J. Mod. Phys. B {\bf 24},  1855 (2010).

\bibitem{belitz94} D.~Belitz and T.R.~Kirkpatrick,
  Rev. Mod. Phys. {\bf 66}, 261 (1994).

\bibitem{pruisken10} A.M.M.Pruisken, in {\it 50 years of Anderson localization},
  ed. by E. Abrahams (World Scientific, 2010), p. 503;
  Int. J. Mod. Phys. B {\bf 24},  1895 (2010).

\bibitem{abrahams01} E.~Abrahams, S.V.~Kravchenko, and M.P.~Sarachik,
  Rev. Mod. Phys. {\bf 73}, 251 (2001);  S.V.~Kravchenko and
  M.P.~Sarachik, Rep. Progr. Phys. {\bf 67}, 1 (2004);
 S.V.~Kravchenko and M.P.~Sarachik, in {\it 50 years of Anderson localization},
  ed. by E. Abrahams (World Scientific, 2010), p. 361;
  Int. J. Mod. Phys. B {\bf 24},  1640 (2010).
  
\bibitem{pudalov04}  V.M.~Pudalov, M.E.~Gershenson, and H.~Kojima,
in \emph{Fundamental Problems of Mesoscopic Physics. Interaction and Decoherence},
ed. by I.V.~Lerner, B.L.~Altshuler, and Y.~Gefen, NATO Sci. Series, Kluwer (2004),
p. 309.

\bibitem{2DMIT}  S.V.~Kravchenko, G.V.~Kravchenko, J.E.~Furneaux, V.M.~Pudalov,
and M.~D'Iorio, Phys. Rev. B \textbf{50}, 8039 (1994); S.V.~Kravchenko, W.E.~Mason,
G.E.~Bowker, J.E.~Furneaux, V.M.~Pudalov, and M.~D'Iorio, Phys. Rev. B \textbf{51},
7038 (1995).

\bibitem{punnoose05} A.~Punnoose and A.M.~Finkelstein, Science {\bf
    310}, 289 (2005).

\bibitem{anissimova07} S.~Anissimova, S.V.~Kravchenko, A.~Punnoose, A.M.~Finkel'stein,
and T.M.~Klapwijk, Nature Phys. \textbf{3}, 707 (2007).

\bibitem{knyazev08} D.A.~Knyazev, O.E.~Omel'yanovskii, V.M.~Pudalov,
  and I.S.~Burmistrov, Phys. Rev. Lett. {\bf 100}, 046405 (2008).

\bibitem{ostrovsky10} P. M. Ostrovsky, I. V. Gornyi, and A. D. Mirlin,
Phys. Rev. Lett. {\bf 105}, 036803 (2010).

\bibitem{feigelman10} M.V. Feigel'man, L.B. Ioffe, V.E. Kravtsov, 
and E. Cuevas, Annals of Physics \textbf{325}, 1368 (2010).

\bibitem{wei88} H.P.~Wei, D.C.~Tsui, M.A.~Paalanen, and
  A.M.M.~Pruisken, Phys. Rev. Lett. {\bf 61}, 1294 (1988).

\bibitem{li05}
W. Li, G. A. Csathy, D. C. Tsui, L. N. Pfeiffer, and K. W. West,
Phys. Rev. Lett. {\bf 94}, 206807 (2005).

\bibitem{waffenschmidt99} S.~Waffenschmidt, C.~Pfleiderer, and
H. v. L\"ohneysen, Phys. Rev. Lett. {\bf 83}, 3005 (1999).

\bibitem{bogdanovich99} S.~Bogdanovich, M.P.~Sarachik, and R.N.~Bhatt,
Phys. Rev. Lett. {\bf 82}, 137 (1999).

\bibitem{LeeWang} D-H.\, Lee and Z.\,Wang,
  Phys. Rev. Lett. \textbf{76}, 4014 (1996).

\bibitem{Wang2} Z.\,Wang, M.P.A.\,Fisher, S.M.\,Girvin, and
  J.T.\,Chalker, Phys. Rev. B \textbf{61}, 8326 (2000).

\bibitem{PB} M.A. Baranov and A.M.M.Pruisken, Europhys. Lett. {\bf
    31}, 543 (1995).

\bibitem{AALR} E. Abrahams, P.W. Anderson, P.A. Lee,
  T.V. Ramakrishnan, Phys. Rev. B {\bf 24}, 6783 (1981).

\bibitem{wegner79} F.~Wegner, Z. Phys. B {\bf 35}, 207 (1979).

\bibitem{PruiskenNLSM} H. Levine, S. Libby, and A.M.M. Pruisken,
  Phys. Rev. Lett. \textbf{51} 1915 (1983); 
A.M.M. Pruisken, Nucl. Phys. B \textbf{235}, 277
  (1984).

\bibitem{PruiskenBurmistrov} A.M.M. Pruisken and I.S. Burmistrov,
  Ann. Phys. (N.Y.) \textbf{316}, 285 (2005).

\bibitem{MirlinReview} A.D. Mirlin, Phys. Rep. {\bf 326}, 259 (2000).

\bibitem{Pruisken85} A.M.M. Pruisken, Phys. Rev. B {\bf 31}, 416 (1985).

\bibitem{Hikami} S. Hikami, Nucl. Phys. B{\bf 215}, 555 (1983).

\bibitem{WegnerB} W. Bernreuther and F.J. Wegner,
  Phys. Rev. Lett. {\bf 57}, 1383 (1986).

\bibitem{Helgason} S. Helgason, \textit{Groups and geometric analysis  
(Integral Geometry, Invariant Differential Operators and Spherical Functions)},
(American Mathematical Society, 2000).

\bibitem{wegner80} F.~Wegner, Z. Physik B {\bf 36}, 209 (1980).

\bibitem{Wegner} D. H\"of, F. Wegner, Nucl. Phys. B {\bf 275}, 561
  (1986); F. Wegner, Nucl. Phys. B {\bf 280}, 193
  (1987); Nucl. Phys. B {\bf 280}, 210 (1987).

\bibitem{Kravtsov}  V.E. Kravtsov and I.V. Lerner, JETP \textbf{61}, 758 (1985);
B.L. Al'tshuler, V.E. Kravtsov, and I.V. Lerner, JETP  \textbf{64}, 1352 (1986)



%\bibitem{PB08} A.M.M. Pruisken and I.S. Burmistrov, JETP Lett. \textbf{87}, 220 (2008).

\bibitem{huckestein95} B.~Huckestein, Rev. Mod. Phys. {\bf 67}, 357
  (1995).

\bibitem{slevin09} K.~Slevin and T.~Ohtsuki,
Phys. Rev. B {\bf 80}, 041304 (2009).

\bibitem{slevin97} K.~Slevin and T.~Ohtsuki,
Phys. Rev. Lett. \textbf{78}, 4083 (1997).

\bibitem{fleishman80} L.~Fleishman and P.W.~Anderson, Phys. Rev. B {\bf
    21}, 2366 (1980); I.V.~Gornyi, A.D.~Mirlin, and D.G.~Polyakov,
  Phys. Rev. Lett. {\bf 95}, 206603 (2005);
D.M.~Basko, I.L.~Aleiner, and B.L.~Altshuler,
  Ann. Phys. (N.Y.) \textbf{321}, 1126 (2006).

\bibitem{polyakov98} D.G.~Polyakov and K.V.~Samokhin,
  Phys. Rev. Lett. {\bf 80}, 1509 (1998).

\bibitem{ludwig08} T.~Ludwig, I.V.~Gornyi, A.D.~Mirlin, and P.~W\"olfle,
Phys. Rev. B {\bf 77}, 235414 (2008).

\bibitem{baranov02} 
M.A.~Baranov,
  A.M.M.~Pruisken, and B.~\v{S}kori\'{c}, Phys. Rev. B {\bf 60}, 16821 (1999);
M.A.~Baranov, I.S.~Burmistrov, and
  A.M.M.~Pruisken, Phys. Rev. B {\bf 66}, 075317 (2002).

\bibitem{thomas82} G.A.~Thomas, Y.~Ootuka, S.~Katsumoto,
  S.~Kobayashi, and W.~Sasaki, Phys. Rev. B {\bf 25},
  4288 (1982);
A. G. Zabrodskii and K.N. Zinov'eva, Sov. Phys. JETP {\bf 59}, 425 (1984);
S.B.~Field and T.F.~Rosenbaum, Phys. Rev. Lett. {\bf 55}, 522 (1985);
M.J.~Hirsch, U.~Thomanschefsky,  and D.F.~Holcomb, Phys. Rev. B {\bf 37},
8257 (1988).

\bibitem{paalanen82}
M.A.~Paalanen, T.F.~Rosenbaum, G.A.~Thomas, and R.N.~Bhatt,
Phys. Rev. Lett. {\bf 48}, 1284 (1982);
G.A.~Thomas, M.A.~Paalanen, and T.F.~Rosenbaum,
Phys. Rev. B {\bf 27}, 3897 (1983);
T.F.~Rosenbaum, R.F.~Milligan, M.A.~Paalanen, G.A.~Thomas, R.N.~Bhatt,
and W.~Lin, Phys. Rev. B {\bf 27}, 7509 (1983).

\bibitem{stupp93}
H.~Stupp, M.~Hornung, M.~Lakner, O.~Madel,  and H.~v.~L\"ohneysen,
Phys. Rev. Lett.{\bf 71}, 2634 (1993);
H.~Stupp, M.~Hornung, M.~Lakner, O.~Madel,  and H.~v.~L\"ohneysen,
Phys. Rev. Lett.{\bf 72}, 2122 (1994).

\bibitem{slevin99} K. Slevin and T. Ohtsuki, 
Phys. Rev. Lett. \textbf{82}, 382 (1999).


\bibitem{koch91}
S.~Koch, R.J.~Haug, K.~v.~Klitzing, and K.~Ploog,
Phys. Rev. Lett. {\bf 67}, 883 (1991).

\bibitem{hohls01}
F.~Hohls, U.~Zeitler, and R.J.~Haug,
Phys. Rev. Lett. {\bf 86}, 5124 (2001);
F.~Hohls, U.~Zeitler, and R.J.~Haug,
Phys. Rev. Lett. {\bf 88}, 036802 (2002).


\bibitem{van-schaijk00}
R. T. F. van Schaijk,  A. de Visser, S.M. Olsthoorn, H.P. Wei, and
A.M.M. Pruisken, Phys. Rev. Lett. {\bf 84}, 1567 (2000); A. de Visser,
L.A. Ponomarenko, G. Galistu, D.T.N. de Lang, A.M.M. Pruisken,
U. Zeitler, and D. Maude,
J. Phys.: Conference Series {\bf 51}, 379 (2006);
A.M.M. Pruisken, D.T.N. de Lang, L.A. Ponomarenko, and A. de
Visser,Sol. State Comm. {\bf 137}, 540 (2006).

\bibitem{karmakar04}
B.~Karmakar, M.R.~Gokhale, A.P.~Shah, B.M.~Arora, D.T.N.~de Lang, A.~de
 Visser, L.A.~Ponomarenko, and A.M.M.~Pruisken, Physica E \textbf{224}, 187 (2004);
L.A. Ponomarenko, D.T.N. de Lang, A. de Visser, V.A. Kubalchinskii, G.B.
Galiev, H. K\"unzel, and A.M.M. Pruisken, Solid State Comm. \textbf{130}, 705 (2004).

\bibitem{li09}
W. Li, C. L. Vicente, J. S. Xia, W. Pan, D. C. Tsui,
L. N. Pfeiffer, and K. W. West,
Phys. Rev. Lett. {\bf 102}, 216801 (2009).

\bibitem{comment} A.M.M. Pruisken and  I.S. Burmistrov, arXiv:0907.0356v1.

%\bibitem{engel93}
%L.W. Engel, D. Shahar, C. Kurdak, and D.C. Tsui,
%Phys. Rev. Lett. {\bf 71}, 2638 (1993).

%\bibitem{hohls02}
%F. Hohls, U. Zeitler, R. J. Haug, R. Meisels, K. Dybko, and F. Kuchar,
%Phys. Rev. Lett. {\bf 89}, 276801 (2002).


\bibitem{Mello} P.A. Mello, J. Phys. A {\bf 23}, 4061 (1990).

\bibitem{Brouwer} P.W. Brouwer and C.W.J. Beenakker,
  J. Math. Phys. {\bf 36}, 4904 (1996).

%\vskip1cm





\end{thebibliography}
\end{document}